\documentclass[letterpaper,11pt]{article}
\usepackage{jheppub}
\usepackage{comment}
\pdfoutput=1
\usepackage{dcolumn}
\usepackage{bm}
\usepackage{graphicx}
\usepackage{diagbox}
\usepackage{amssymb,amsmath}
\usepackage{multirow}
\usepackage{color,url}
\usepackage{tabu}
\usepackage{array}
\usepackage{ulem}
\usepackage{soul}
\usepackage{colordvi}
\usepackage{subcaption}
\usepackage{ mathrsfs }
\def\ga{\mathrel{\raise.3ex\hbox{$>$\kern-.75em\lower1ex\hbox{$\sim$}}}}
\def\la{\mathrel{\raise.3ex\hbox{$<$\kern-.75em\lower1ex\hbox{$\sim$}}}}
\def\beqa{\begin{eqnarray}}
\def\eeqa{\end{eqnarray}}

\begin{document}

\title{\boldmath  Exploring the Universality of Hadronic Jet Classification}

\author[1,5]{Kingman Cheung,}
\author[1]{Yi-Lun Chung,}
\author[2]{Shih-Chieh Hsu,}
\author[3,4]{and Benjamin Nachman}
\affiliation[1]{\normalsize Department of Physics and Center for Theory and Computation, National Tsing Hua University, Hsinchu 300, Taiwan}
\affiliation[2]{\normalsize Department of Physics, University of Washington, Seattle, Washington 98195, USA}
\affiliation[3]{\normalsize Physics Division, Lawrence Berkeley National Laboratory, Berkeley, CA 94720, USA}
\affiliation[4]{\normalsize Berkeley Institute for Data Science, University of California, Berkeley, CA 94720, USA}
\affiliation[5]{\normalsize Division of Quantum Phases and Devices, 
School of Physics,Konkuk University, Seoul 143-701, Republic of Korea}

\emailAdd{cheung@phys.nthu.edu.tw}
\emailAdd{s107022801@m107.nthu.edu.tw}
\emailAdd{schsu@uw.edu}
\emailAdd{bpnachman@lbl.gov}

\date{\today}

\abstract{
The modeling of jet substructure significantly differs between Parton Shower Monte Carlo (PSMC) programs.  Despite this, we observe that machine learning classifiers trained on different PSMCs learn nearly the same function.  This means that when these classifiers are applied to the same PSMC for testing, they result in nearly the same performance.  This classifier universality indicates that a machine learning model trained on one simulation and tested on another simulation (or data) will likely be optimal.  Our observations are based on detailed studies of shallow and deep neural networks applied to simulated Lorentz boosted Higgs jet tagging at the LHC.
 
}

\maketitle

\newpage
\section{Introduction}\label{introduction}

Deep learning is becoming widely used for various classification tasks in collider physics (see e.g., Ref.~\cite{Feickert:2021ajf,Larkoski:2017jix,Guest:2018yhq,Radovic:2018dip,Bourilkov:2019yoi,Karagiorgi:2021ngt}).  One of the core benefits of deep learning over traditional analysis techniques is that it is able to identify patterns in very high-dimensional feature spaces.  At the Large Hadron Collider (LHC), such low-level inputs are dominated by hadronic activity.  Most machine learning approaches are trained using Parton Shower Monte Carlo (PSMC) simulations that produce exclusive final states with the same complexity as real data~\cite{Buckley:2011ms}.  However, there are significant variations between PSMCs due to the large number of perturbative and non-perturbative modeling assumptions.

These variations lead to potential biases and suboptimal sensitivity in data analyses~\cite{Nachman:2019dol}.  A bias occurs when the simulation model used for inference (given an analysis strategy) is not the same as nature.  There is a large and growing literature on methods to reduce biases from PSMC model variations through decorrelation~\cite{Louppe:2016ylz,Englert:2018cfo,Wunsch:2019qbo,clavijo2020adversarial} and other approaches~\cite{deCastro:2018mgh,Ghosh:2021roe,Simpson:2022suz}. A key challenge with modeling uncertainties in contrast to experimental uncertainties is that they are often estimated by comparing two simulations.  This difference does not have a statistical origin and may not be the full uncertainty, so caution is required to reduce the uncertainty through automated approaches~\cite{Ghosh:2021hrh}.  A general solution to estimating (and then reducing) systematic uncertainties from PSMC variations is still an active area of research and development\footnote{See Ref.~\cite{Nachman:2019yfl,Stein:2022nvf} for the possibility of using machine learning to bound these uncertainties.}.

In principle, the same challenge exists when quantifying suboptimal performance due to PSMC variations.  Suboptimal performance occurs when the simulation model used for training a machine learning model is different than nature.  While not directly a source of systematic uncertainty, this suboptimality has important consequences for the physics program of the LHC.  To quantify the suboptimality, one could compare different PSMC models, as is done for determine the systematic uncertainty.  This has the same unsatisfying properties as described above.  

However, there have been a number of hints in the literature that the suboptimality due to PSMC variations may actually be small.  For example, Ref.~\cite{Komiske:2016rsd} observed that training a quark versus gluon jet classifier with the \textsc{Herwig}~\citep{Bellm:2015jjp} PSMC and then applying it to jets simulated with the \textsc{Pythia}~\citep{Sjostrand:2007gs} PSMC has nearly the same performance as training with \textsc{Pythia} and also testing on \textsc{Pythia} (with a statistically identical, but independent dataset).  This small difference in performance is contrasted to the large difference in performance when testing on jets from \textsc{Herwig}.  From this observation, we conjecture that the deep learning models are learning universal properties of quantum chromodynamics (QCD).  We hypothesis that the performance gaps present when the test sets differ simply reflects variations in the amount of QCD radiation, but not the type of information that is useful for discrimination. 

To build intuition for this conjecture, consider the case of quark versus gluon jet tagging.  At leading logarithmic (LL) order and considering only infrared and collinear safe observables, the optimal classifier is simply the number of perturbative emissions inside the jet~\cite{Frye:2017yrw}.  This statement is true independent of the strong coupling constant, $\alpha_s$. However, common metrics of performance such as the Area Under the Curve (AUC) depend on $\alpha_s$; when there are more emissions (higher $\alpha_s$), the quark and gluon perturbative multiplicity distributions are more separable.  In particular, at LL, perturbative multiplicity is a Poisson random variable with a mean that is proportional to a color factor multiplied by $\alpha_s$.  As $\alpha_s$ grows, the gluon distribution grows significantly faster than the quark one:
\begin{align}
    \frac{\mu_g-\mu_q}{\sqrt{\sigma_g^2+\sigma_q^2}}&\sim \frac{\alpha_s(C_F-C_A)}{\sqrt{\alpha_s C_F+\alpha_s C_A}}\propto\sqrt{\alpha_s}\,,
\end{align}
where $C_F=4/3$ ($C_A=3$) is the quark (gluon) color factor.
Imagine that two PSMCs had the same physics approximations, but different values of $\alpha_s$.  They would find the same classifier and thus if the test set is the same, the performance would be the same.

Our goal is to test the universality hypothesis in detail using the important benchmark problem of Lorentz boosted Higgs boson jet versus QCD jet tagging.  We consider both shallow and deep learning models as well as a variety of PSMC models.

This paper is organized as follows.  A concrete example are introduced in Sec. \ref{Numerical Examples}. Architectures of  deep-learning classifiers are in Sec. \ref{Architecture of Classifiers}. The results are provided in Sec. \ref{Results}. The paper ends with conclusions and outlook in Sec. \ref{Conclusions and Outlook}.

\section{Numerical Examples}\label{Numerical Examples}
Lorenz-boosted Higgs tagging, focusing on the $b\bar{b}$ final state, is the example in this study. High-level features and low-level inputs are used to train shallow and deep-learning classifiers.

\subsection{Monte Carlo Samples} \label{mc_samples}
This study considers Lorenz-boosted Higgs tagging, focusing on the $b\bar{b}$ final state.  The signal is high $p_T$ Higgs bosons and the background is generic quark and gluon jets. The hard-scatter reactions are common to all parton shower models and are generated with MadGraph5\_aMC@NLO 2.7.3~\citep{Alwall:2014hca} for modeling $pp$ collisions at $\sqrt{s}$ = 14 TeV. The {\texttt{PDF4LHC15\_nnlo\_mc}} \citep{Butterworth:2015oua} parton distribution function and the {\texttt{NNPDF30\_nlo\_as\_0118}} \citep{NNPDF:2014otw} parton distribution function are used for signal and background, respectively. 

The hard-scattering events are passed to \textsc{Pythia 8.303}~\citep{Sjostrand:2007gs} to simulate the parton shower and hadronization, using three different complete parton-shower frameworks. 
The first one is default setting, where evolution variable is virtuality of the off-shell propagator. The second framework is Virtual Numerical Collider with Interleaved Antennae (\textsc{Vincia}) shower\citep{Skands:2020lkd, Brooks:2020mab, Fischer:2016vfv}, where the evolution variable is transverse momentum for QCD + EW/QED showers based on the antenna formalism. The last framework is Dipole resummation (\textsc{Dire}), which is a transverse-momentum ordered dipole shower. \textsc{Herwig} 7.2.2~\citep{Bellm:2015jjp} with angularly-ordered showers is also used to model the parton shower and hadronization. \textsc{Pyjet}~\citep{noel_dawe_2021_4446849,Cacciari:2011ma} and the anti-$k_t$~\citep{Cacciari:2008gp} algorithm with radius parameter $R$ = 1.0 are used to define the jets.

An event preselection similar to Ref.~\cite{Lin:2018cin} is used to reject most background events.  The Higgs-like jet is required to satisfy 300 GeV $<$ $p^J_T$ $<$ 500 GeV, 110 GeV $<$ invariant mass of the jet ($M_J$) $<$ 160 GeV and to be double $b$-tagged. Jets are declared double $b$-tagged if they have two or more ghosted-associated~\citep{Cacciari:2008gn,Buckley:2015gua} $B$ hadrons. After the preselection, the high-level jet features and low-level features are used to probe the universality of discriminating boosted Higgs jets from QCD jets.

Since the goal of this paper is to investigate the universality of hadronic jet classification, there are a number of simplifying assumptions. The background in the study is only generic quark and gluon jets. The relatively smaller $t\bar{t}$ background is ignored. For each PSMC setup, the default parameters are used. 

\subsection{High-level Features}\label{High-level Features}

In order to distinguish Higgs jets via Gradient Tree Boosting (BDT) and a fully connected / dense neural network, the following six commonly-used high-level features are considered:\\

\quad
1. $M_J$ : invariant mass of the leading jet;\\

\quad
2. $\tau_{21}=\tau_2/\tau_1$ : $n$-subjettiness ratio~\cite{Thaler:2010tr,Thaler:2011gf};\\
 
\quad
3. $D_2^{(\beta)}=e_3^{(\beta)}/(e_2^{(\beta)})^3$ with $\beta=1,2$ : energy correlation function ratios~\cite{Larkoski:2014gra};\\

\quad
4. $C_2^{(\beta)}=e_3^{(\beta)}/(e_2^{(\beta)})^2$ with $\beta=1,2$ :  energy correlation function ratios~\cite{Larkoski:2013eya};\\

\noindent  where $e_i$ is the normalized sum over doublets ($i=2$) or triplets ($i=3$) of constituents inside jets, weighted by the product of the constituent transverse momenta and pairwise angular distances. For this analysis, $\beta$ is considered to be 1 and 2.

\begin{figure}[h]
\centering
     \begin{subfigure}{0.45\textwidth}
        \centering
        \includegraphics[width=2.5in]{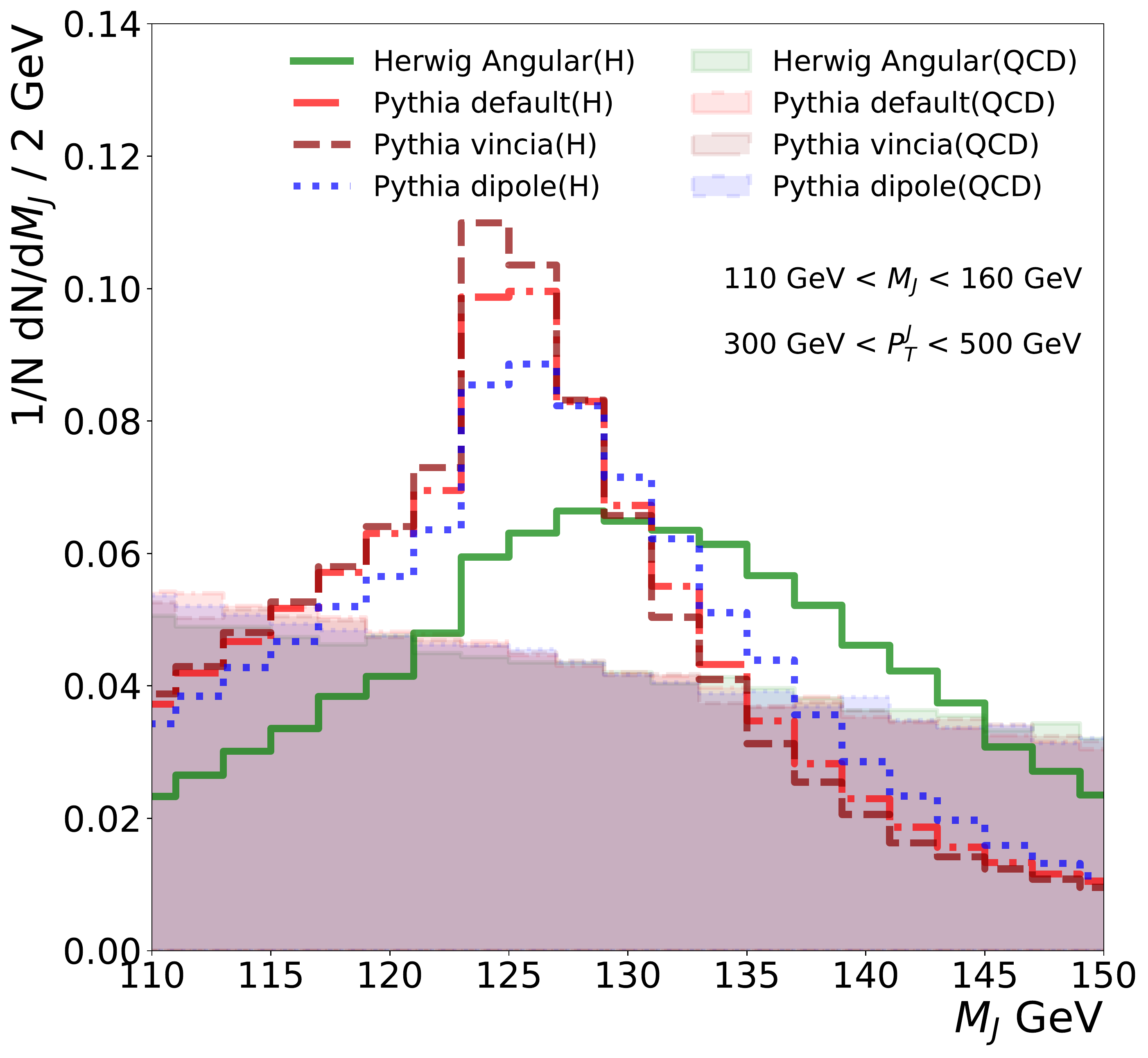}
     \end{subfigure}
     \begin{subfigure}{0.45\textwidth}
        \centering
        \includegraphics[width=2.5in]{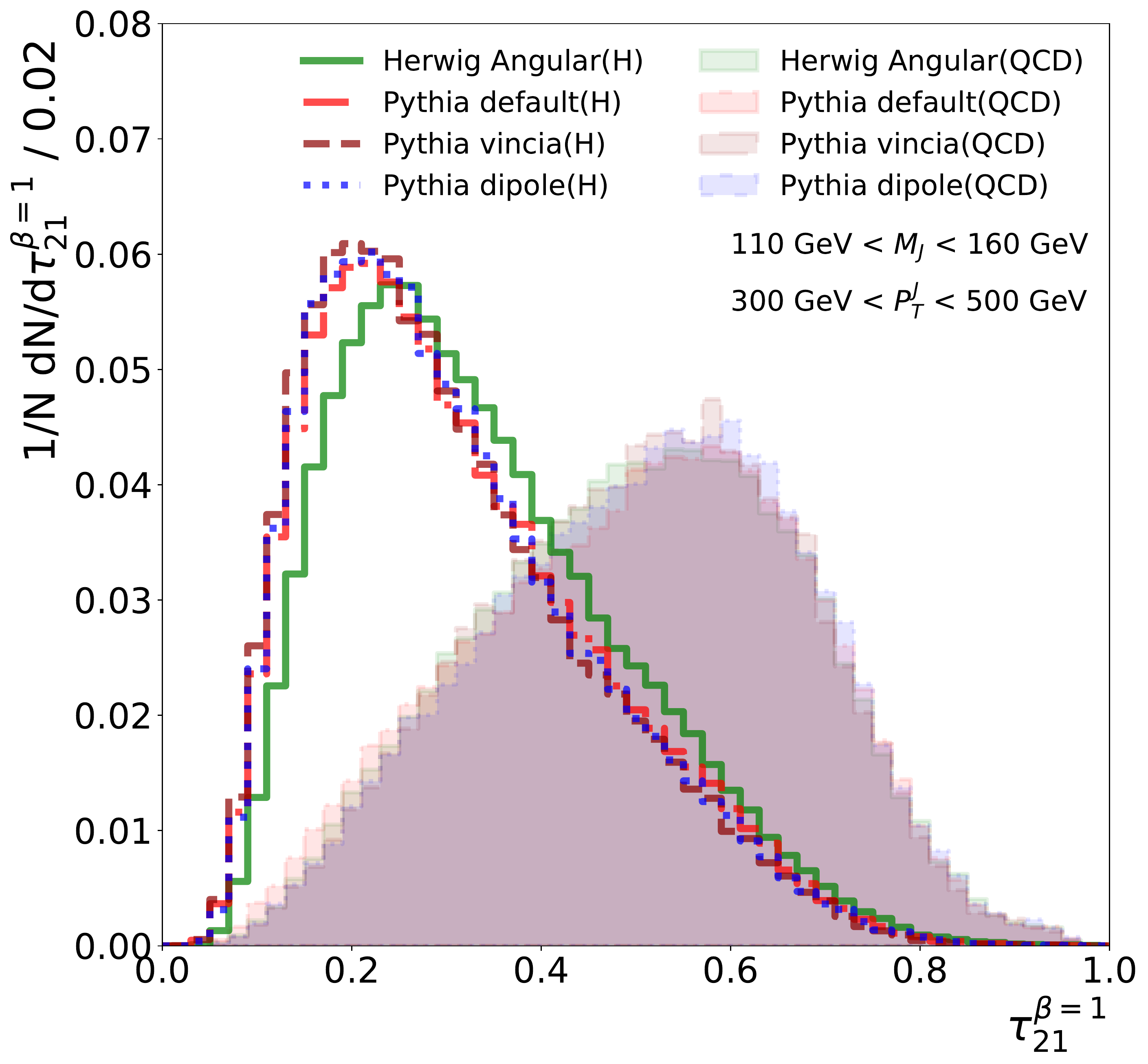}
     \end{subfigure}
     \begin{subfigure}{0.45\textwidth}
        \centering
        \includegraphics[width=2.5in]{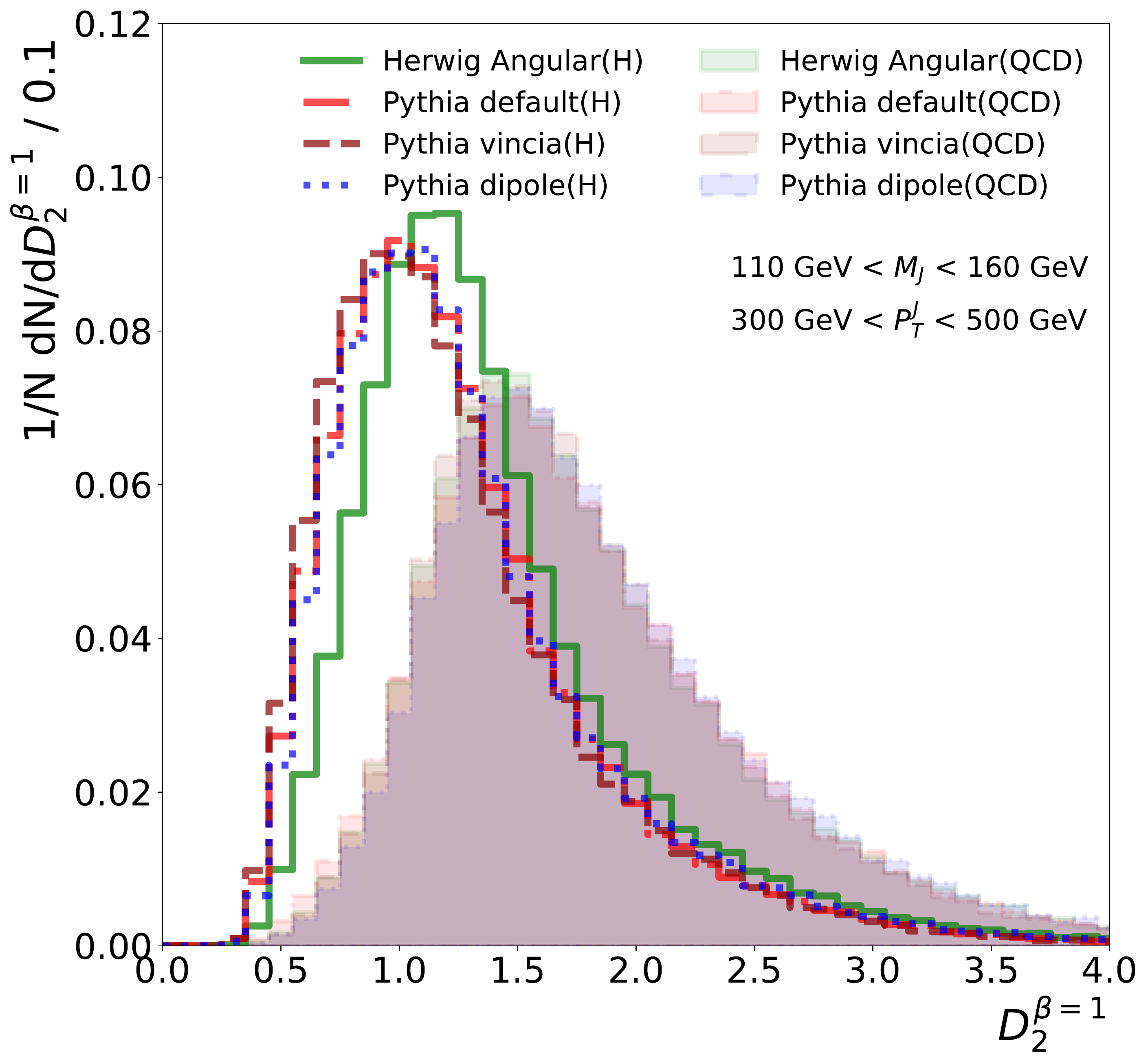}
     \end{subfigure}
     \begin{subfigure}{0.45\textwidth}
        \centering
        \includegraphics[width=2.5in]{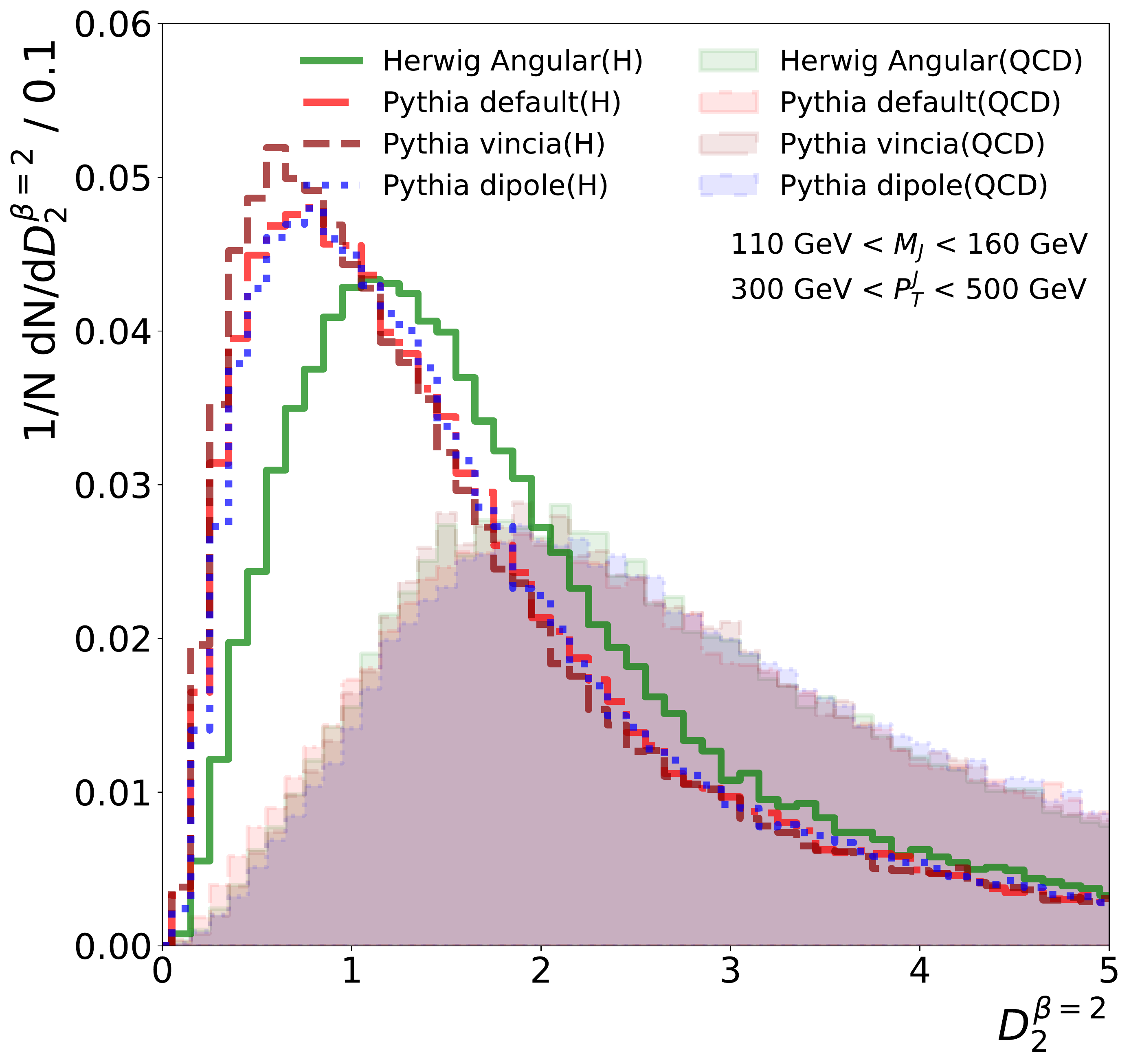}
     \end{subfigure}
          \begin{subfigure}{0.45\textwidth}
        \centering
        \includegraphics[width=2.5in]{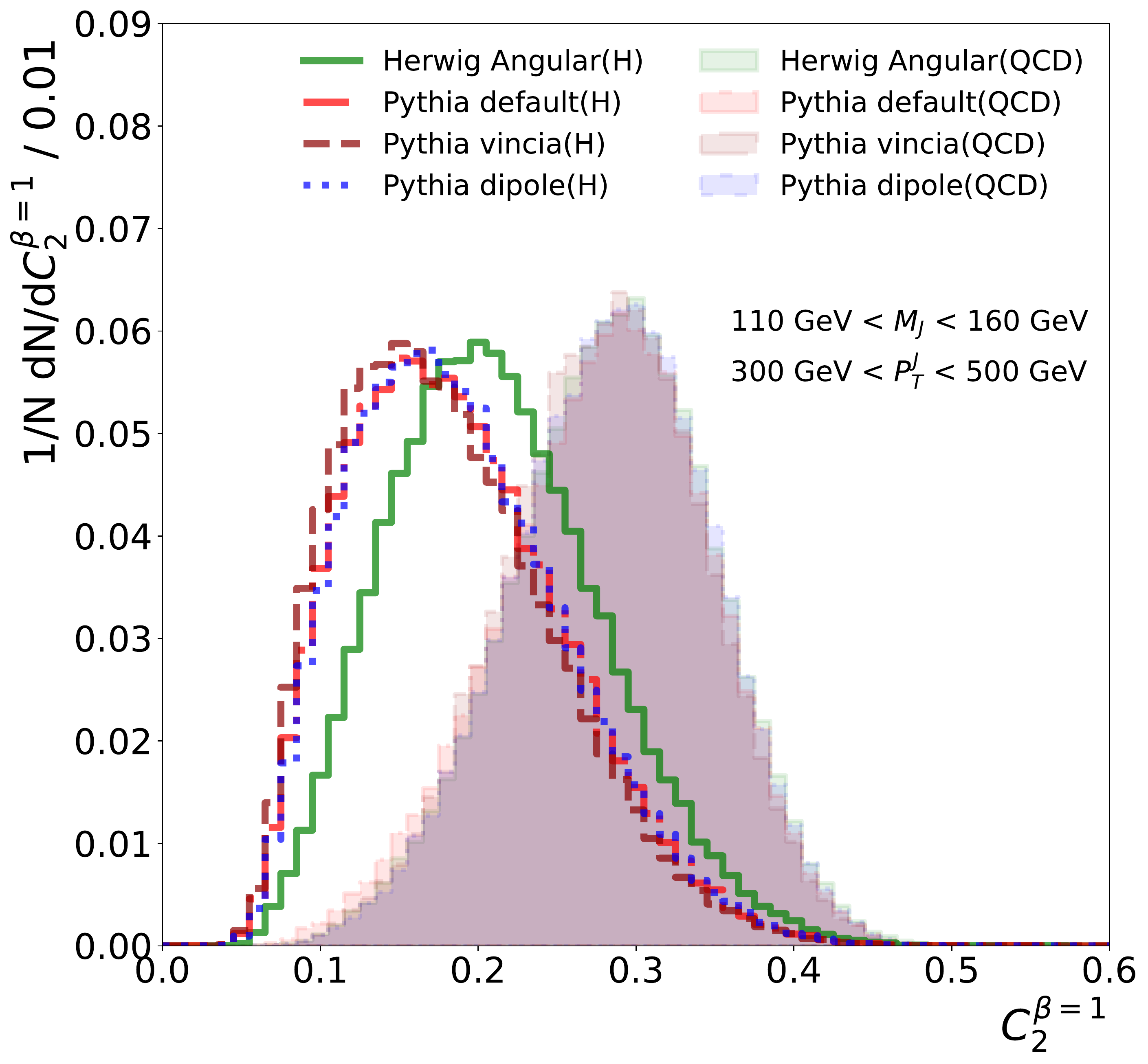}
     \end{subfigure}
          \begin{subfigure}{0.45\textwidth}
        \centering
        \includegraphics[width=2.5in]{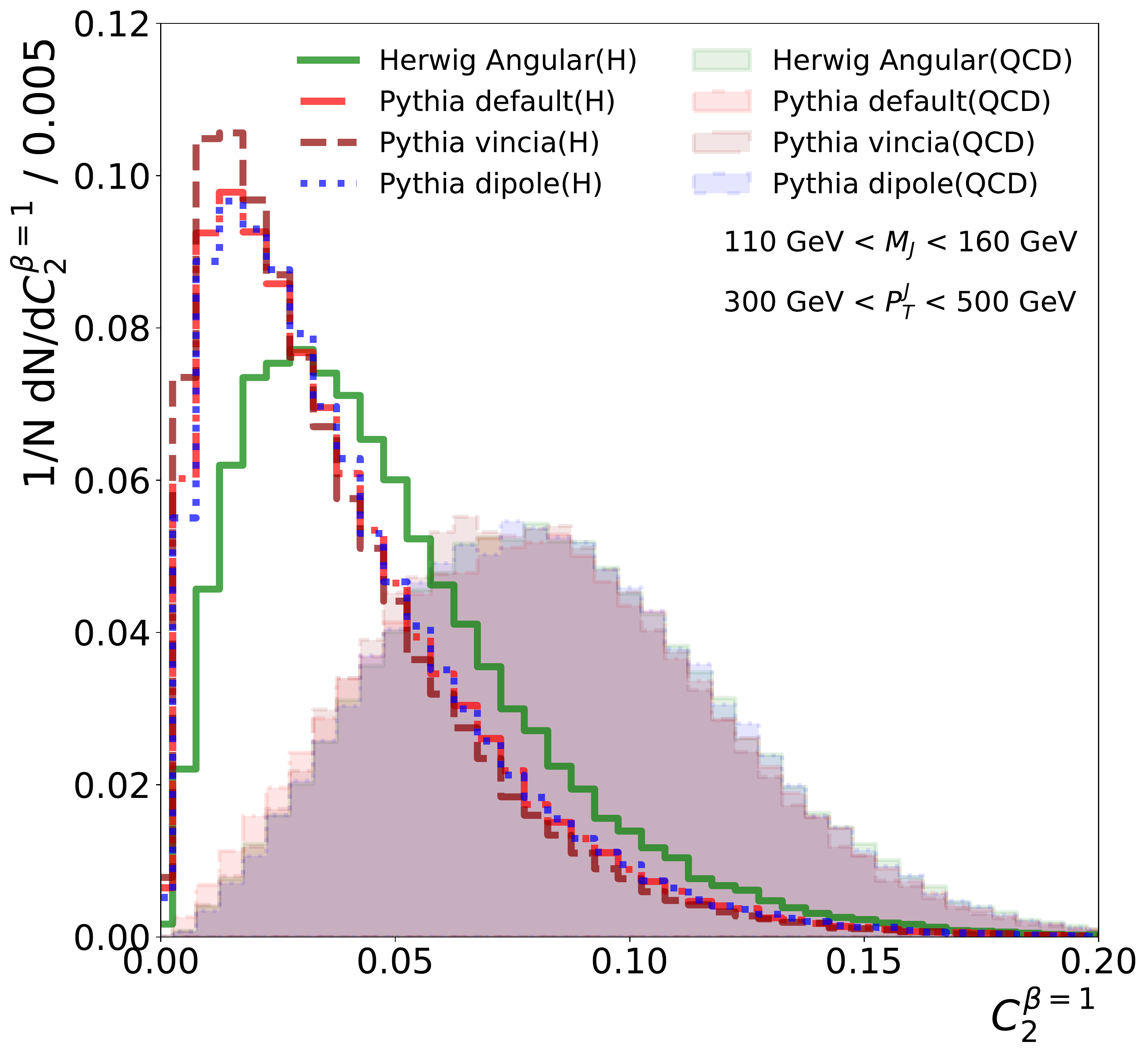}
     \end{subfigure}
\caption{The six high-level features used to distinguish boosted Higgs boson jets from QCD jets events.}
\label{fig:high_level_features}
\end{figure}


The distributions of these six variables are shown in Fig. \ref{fig:high_level_features}, in which the capability of each observable to discriminate between signal and background is demonstrated. The salient features of these histograms are described below.

The jet invariant mass distribution peaks near the Higgs boson mass of 125 GeV~\cite{10.1093/ptep/ptaa104} for the signal and has a broad distribution for the background. In the setup of this study, \textsc{Herwig} 7.2.2 with angularly-ordered showers leads to slightly higher and broader signal peak due to different underlying event structure compared to \textsc{Pythia 8.303}. 
Similarly, the distributions of $\tau_{21}$, $D_2^\beta$, and $C_2^\beta$ show similar position and shape of the peak among the \textsc{Pythia} PSMC's, but somewhat different for the  \textsc{Herwig Angular}. 
The two-prong structure due to the decay of massive objects into two hard QCD partons in the case of the signal jets results in low $\tau_{21}$, $D_2$ and $C_2$. 

\subsection{Low-level Features}\label{Low-level Features}

\begin{figure}[h]
\centering
     \begin{subfigure}{0.45\textwidth}
        \centering
        \includegraphics[width=2.8in]{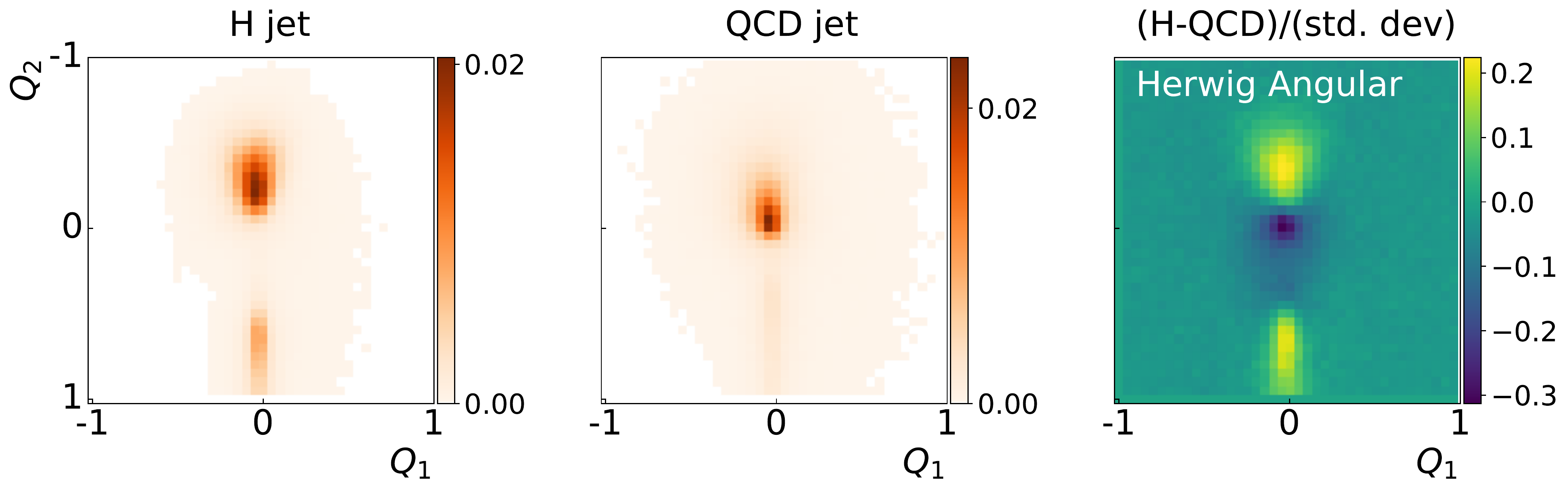}
        \caption{\textsc{Herwig} angular}
     \end{subfigure}
     \begin{subfigure}{0.45\textwidth}
        \centering
        \includegraphics[width=2.8in]{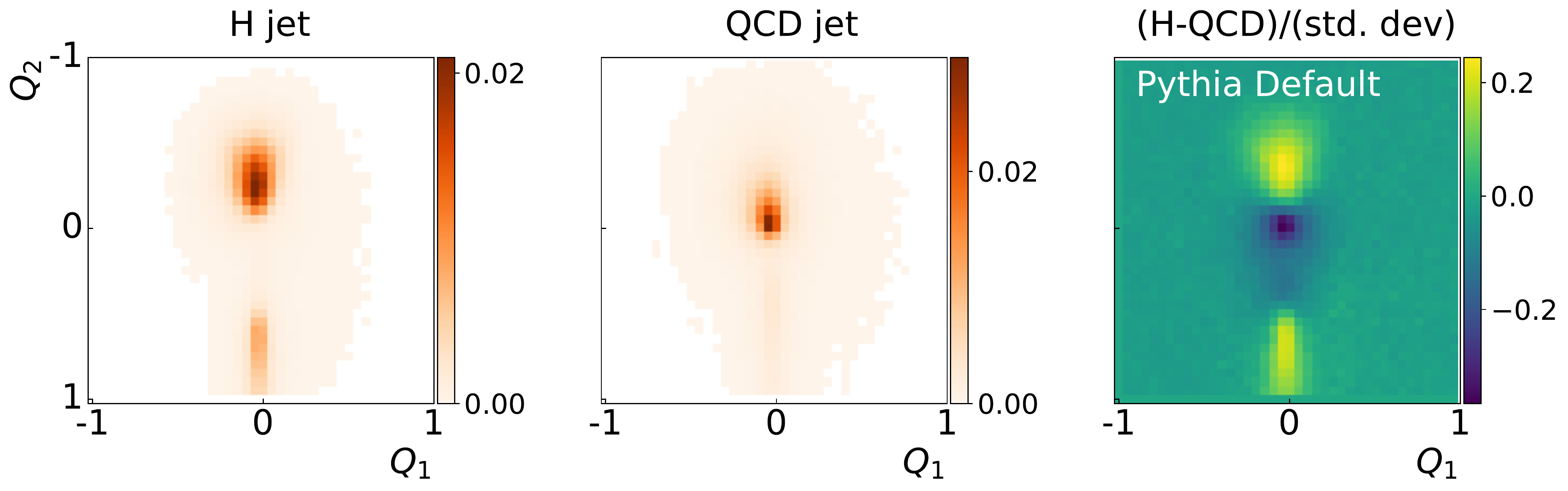}
        \caption{\textsc{Pythia} Default}
     \end{subfigure}
     \begin{subfigure}{0.45\textwidth}
        \centering
        \includegraphics[width=2.8in]{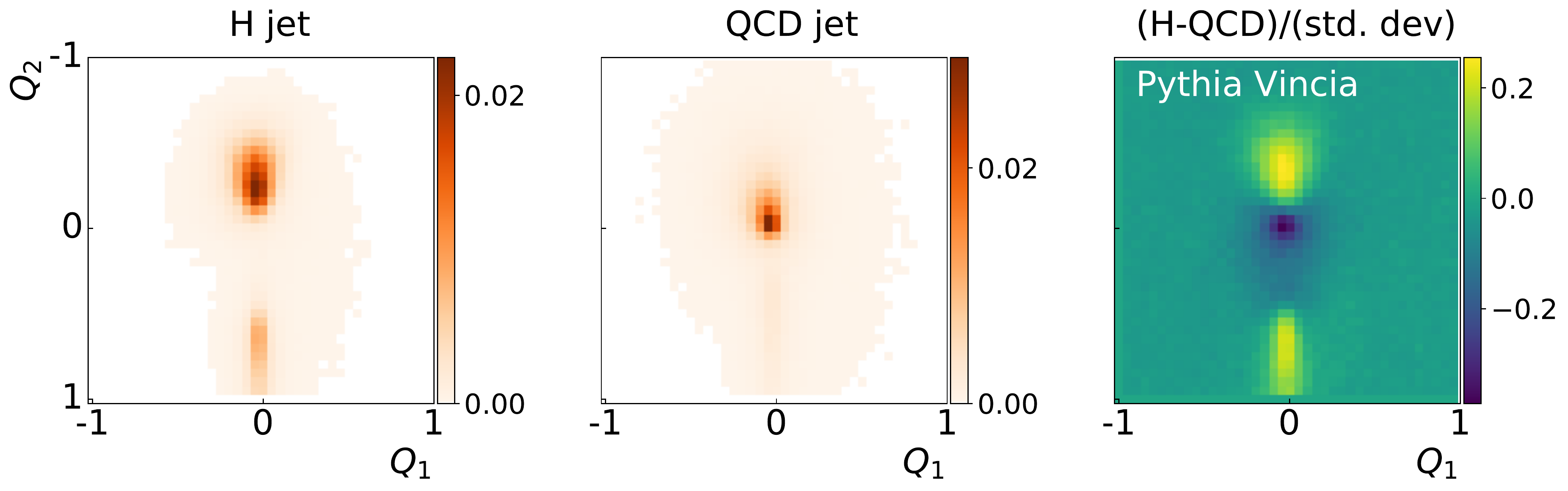}
        \caption{\textsc{Pythia} VINCIA}
     \end{subfigure}
     \begin{subfigure}{0.45\textwidth}
        \centering
        \includegraphics[width=2.8in]{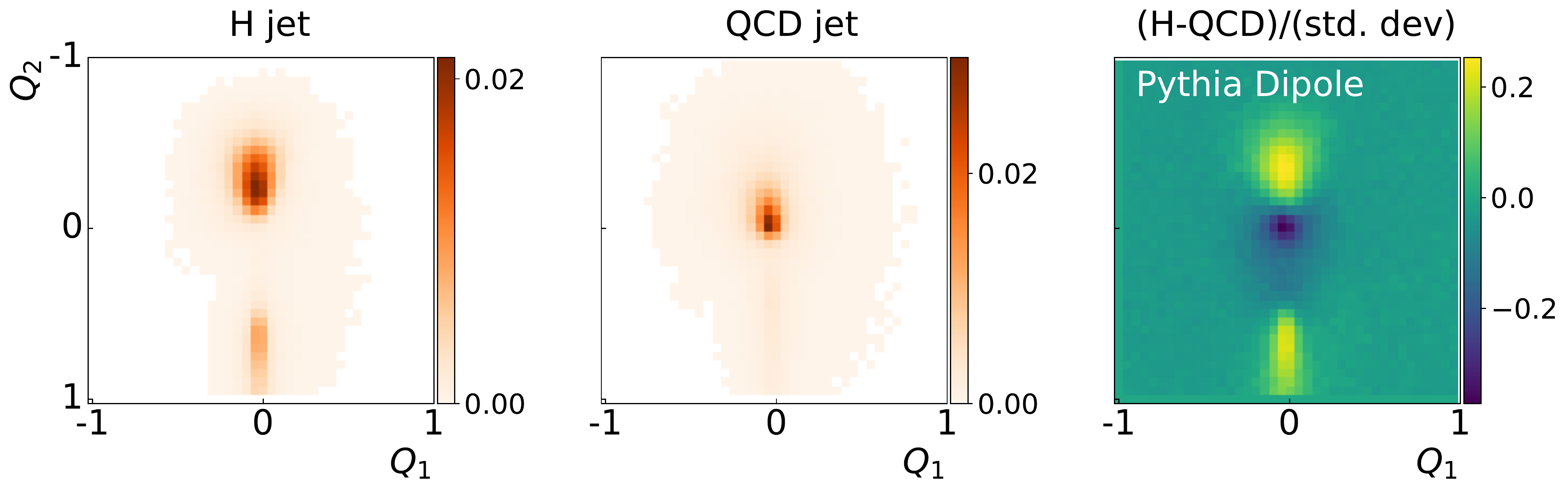}
        \caption{\textsc{Pythia} Dipole}
     \end{subfigure}
\caption{Low-level features. The average of 40000 Higgs-like jet images in the charged $p_T$ channel (left column and middle column). $Q_1$ and $Q_2$ denote the new axes after the jet's axis is centralized and rotated. The intensity in each pixel is the sum of the charged particle $p_T$. The total intensity in each image is normalized to unity.  Images in right column are the average difference between Higgs jet and QCD jet images.}
\label{fig:jet_images}
\end{figure}

The low-level inputs to the CNN are images of Higgs-like jet~\cite{Cogan:2014oua,deOliveira:2015xxd}. The resolution is 40$\times$40 pixels and in 1R$\times$1R range, where $R$ is the jet radius. The images consist of three channels, analogous to the Red-Green-Blue (RGB) channels of a color image~\cite{Komiske:2016rsd}. The pixel intensity for the three channels correspond to the sum of the charged particle $p_T$, the sum of the neutral particle $p_T$, and the number of charged particles in a given region of the image. The Higgs-like jet images are rotated to align along two-subject's axis. The leading subjet is at the origin and the subleading subjet  is directly below the leading subjet. If there is a third-leading subjet, the image will be reflected.
All images are normalized so that the intensities all sum to unity\footnote{This may remove useful discriminating information; however, it significantly improves the stability of the machine learning training~\cite{deOliveira:2017pjk}.}.  After normalization, the pixel intensities are standardized so that their distribution has mean zero and unit variance.  Figure \ref{fig:jet_images} shows the average Higgs-like jet images in the charged $p_T$ channel. The patterns in the charged $p_T$ channel are similar to the other two channels.

\begin{figure}[h]
\centering
     \begin{subfigure}{0.99\textwidth}
        \centering
        \includegraphics[width=6in]{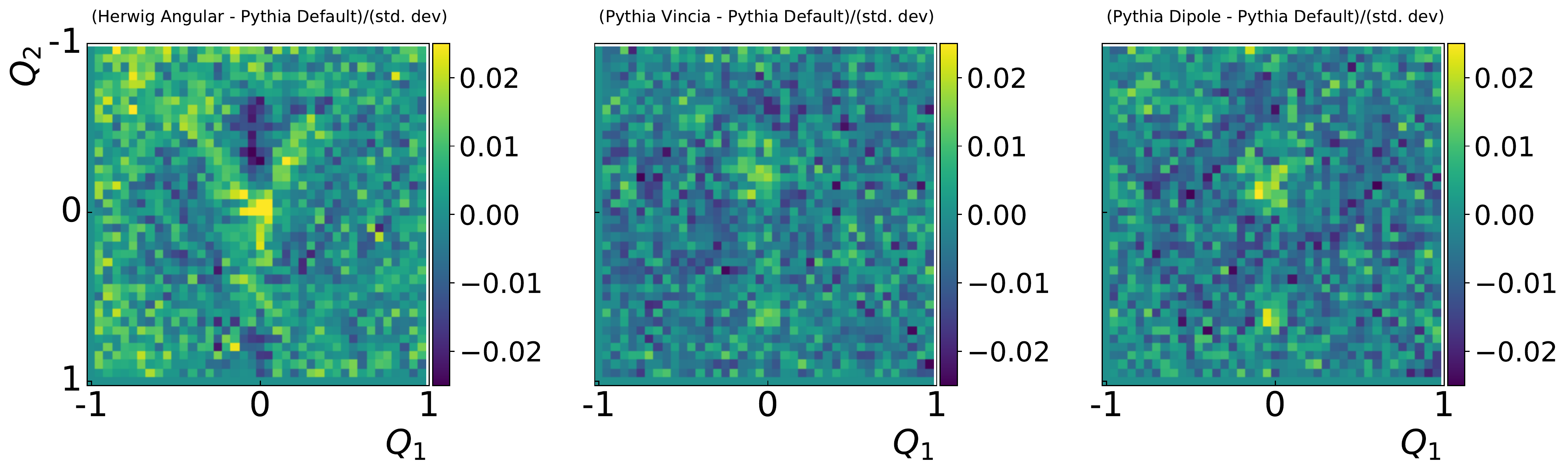}
        \caption{Higgs jet images comparison to \textsc{Pythia} Default samples.}
     \end{subfigure}
     \begin{subfigure}{0.99\textwidth}
        \centering
        \includegraphics[width=6in]{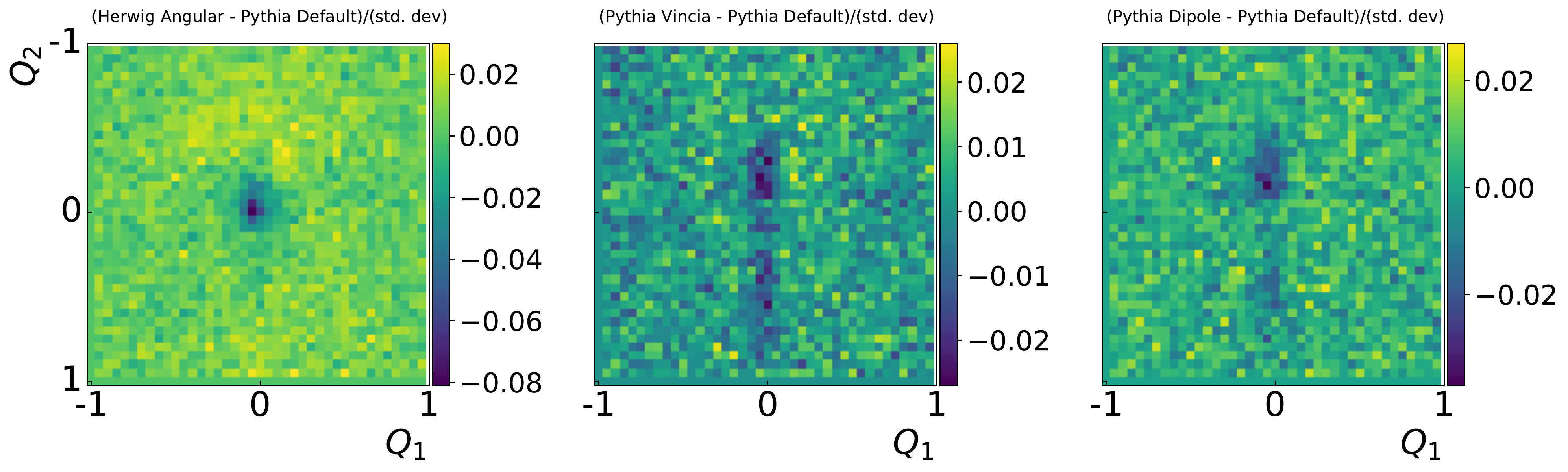}
        \caption{QCD jet images comparison to \textsc{Pythia} Default samples.}
     \end{subfigure}
\caption{The average difference between the other generators and the \textsc{Pythia} Default showering (a) Higgs-like jet images, and (b) QCD jet images. $Q_1$ and $Q_2$ denote the new axes after the jet's axis is centralized and rotated.}
\label{fig:jet_images_comparison}
\end{figure}

Figure~\ref{fig:jet_images_comparison} shows the difference between the four PSMC algorithms with respect to \textsc{Pythia 8.303} default showering, referred to as the nominal simulation. The substructure in jets are different among the other three PSMC simulations with respect to the nominal sample due to different approximations made in the final state radiation and other QCD effects. This diversity of the PSMC approaches may effect the performance of jet classifiers trained on low-level features. Therefore, we train a convolutional neural network-based jet classifier to explore this generator-dependence of classification performance.

\section{Classifier Architectures }\label{Architecture of Classifiers}

The BDT has a fixed number of estimators (1400) with maximum depth 5. The minimum number of samples is fixed at 5\% as required to split an internal node and 1\% as required to be at a leaf node. 
This BDT model is trained on the high-level features of the jet using the {\texttt{scikit-learn}} library \cite{scikit-learn}. KerasTuner~\cite{omalley2019kerastuner} is used to get the best configuration of hyperparameters.

The dense neural network has four full connected layers. There are 224, 928, 288 and 1024 neurons, respectively. Rectified linear unit (ReLU) activation functions are used for all layers of this neural network. Before the output layer, Dropout~\citep{JMLR:v15:srivastava14a} regularization is added to reduce overfitting with a dropout rate = 0.01. For this two-class problem, the activation function of the output layer is a sigmoid function. The binary cross entropy loss function is optimized during the training. The Adam optimizer~\citep{kingma2017adam} with a learning rate of 6.5428$\times10^{-5}$ is used to select the network weights. The KerasTuner~\cite{omalley2019kerastuner} is used to get the best configuration of hyperparameters. The {\texttt{Keras-2.4.0}} library is used to train the dense neural network models with the {\texttt{T{\footnotesize{ENSORFLOW}}-2.4.1}} \cite{tensorflow2015-whitepaper} backend, on a {\texttt{NVIDIA A100 SXM 80GB}} Graphical Processing Unit (GPU).

Details of the CNN are as follows. The convolution filter is 5$\times$5, the maximum pooling layers are 2$\times$2, and the stride length is 1. ReLU activation functions are used for all intermediate layers of the neural network. The first convolution layer has 96 filters and the second convolution layer in each stream has 32 filters. A flatten layer is used after the second maximum pooling layer. Two dense layers are connected to the flatten layer with 350 and 400 neurons, respectively. Before the output layer, Dropout regularization is added with a dropout rate = 0.01. As for the dense network, the last activation is a sigmoid function and binary cross entropy is optimized during training. The AdaDelta optimizer~\citep{DBLP:journals/corr/abs-1212-5701} with learning rate 6.0216$\times10^{-3}$ is used to select the network weights. The KerasTuner~\cite{omalley2019kerastuner} is used to get the best configuration of hyperparameters. The same setup as for the dense network is used to run the CNN.

\section{Results}\label{Results}

In this study, the receiver operating characteristic curve (ROC), the area under the ROC curve (AUC), the maximum significance improvement characteristic (SIC) and rejection (inverse background efficiency) at 50\% signal efficiency are used to be metrics to quantify the universality.  The AUC is between 0.5 (poor classification performance) and 1 (maximum classification performance).  The SIC is the signal efficiency divided by the square root of the background efficiency and represents by how much (as a multiplicative factor) the significance would improve with a given threshold on the classifier score.  The maximum SIC is simply the maximum SIC attained across all thresholds.  In order to quantify the variation from classifier training itself, the performance is evaluated by $k$-fold cross-validation technique with $k=50$.  In this procedure, the datasets are randomly partitioned into 50 parts and for each one, the other 49 sets are used for constructing the classifier.  The mean and spread over the folds is used to quantify the model performance. 

\begin{figure}[h]
\centering
     \begin{subfigure}{0.3\textwidth}
        \centering
        \includegraphics[width=2.2in]{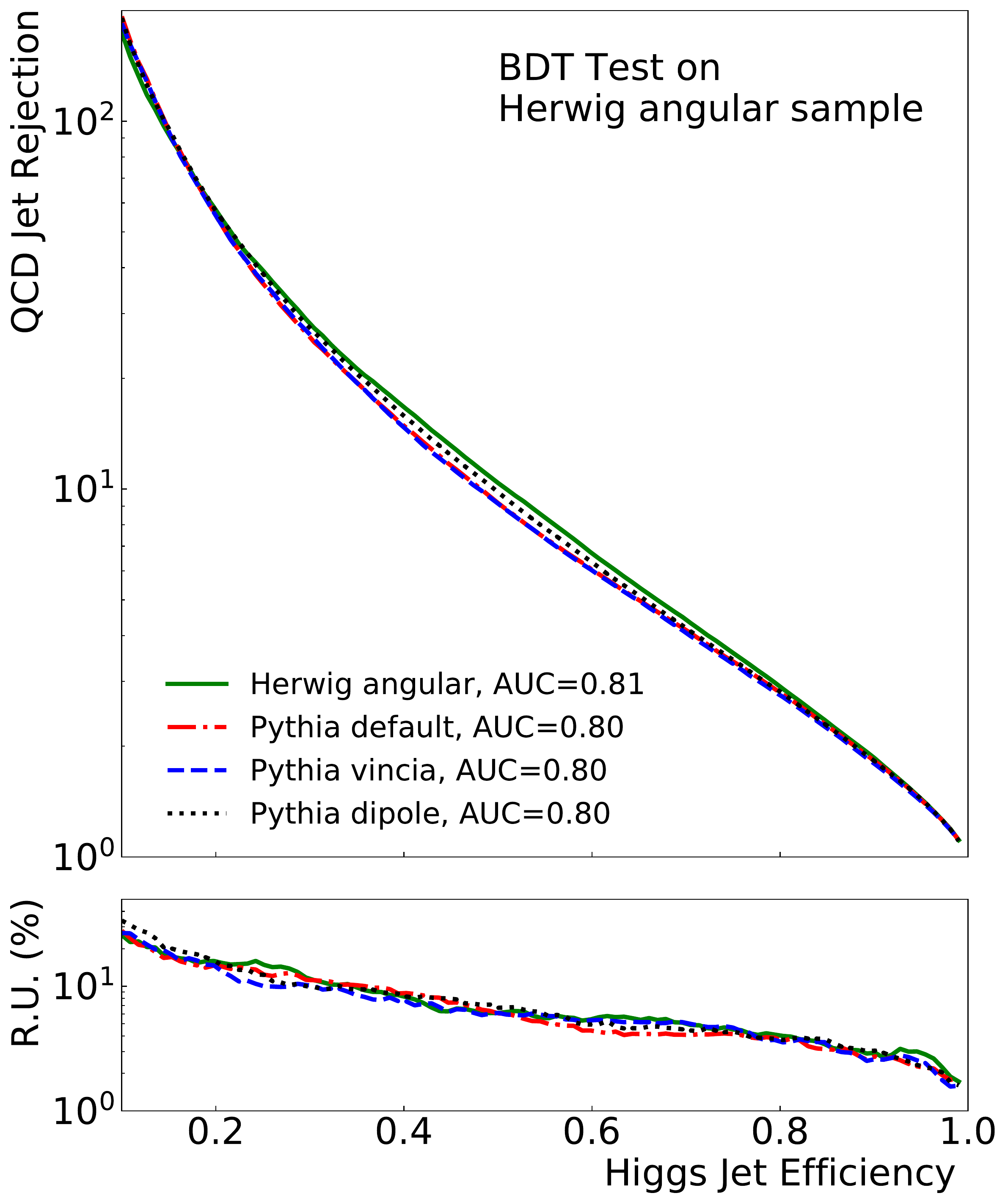}
     \end{subfigure}
     \begin{subfigure}{0.3\textwidth}
        \centering
        \includegraphics[width=2.2in]{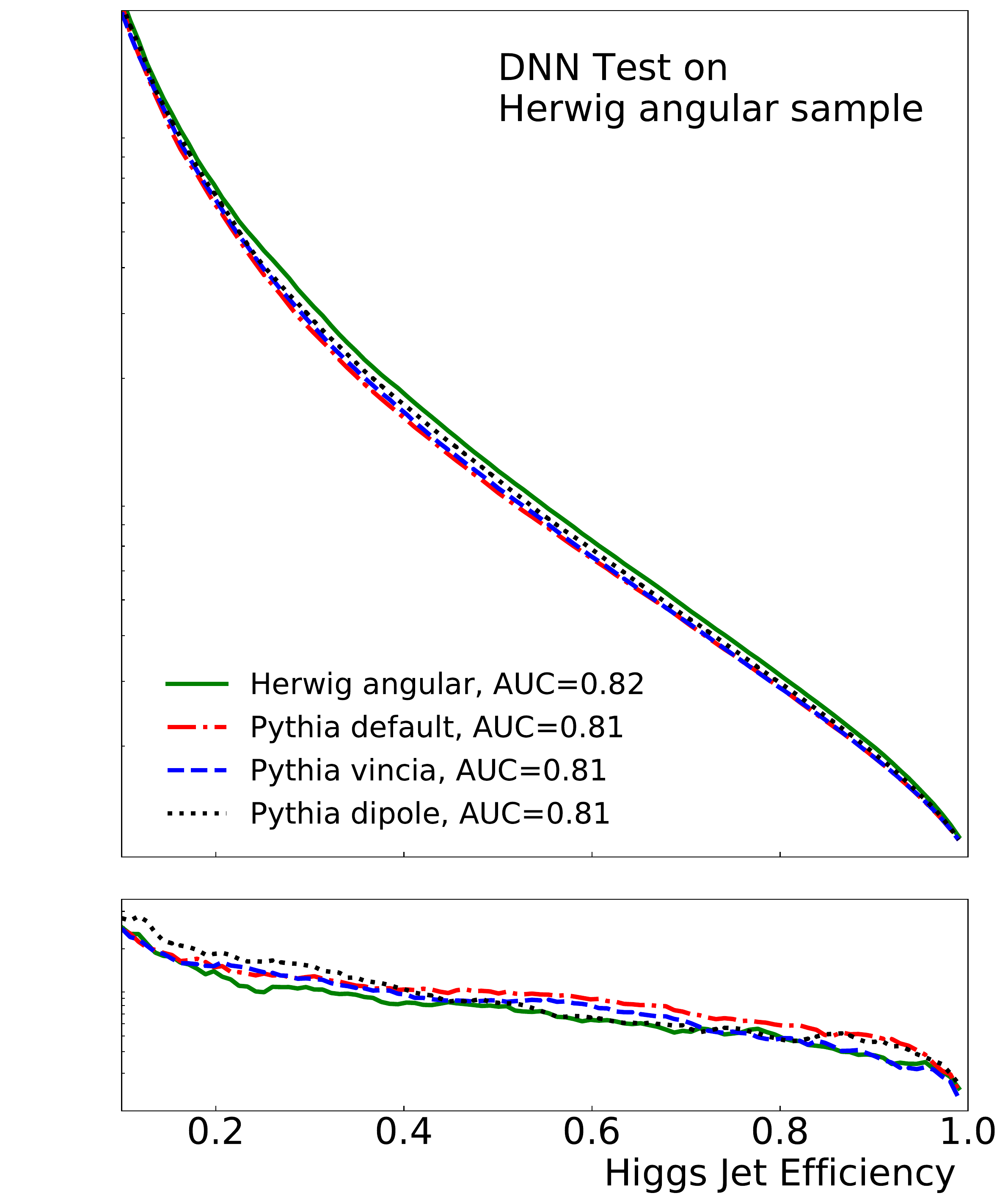}
     \end{subfigure}
     \begin{subfigure}{0.3\textwidth}
        \centering
        \includegraphics[width=2.2in]{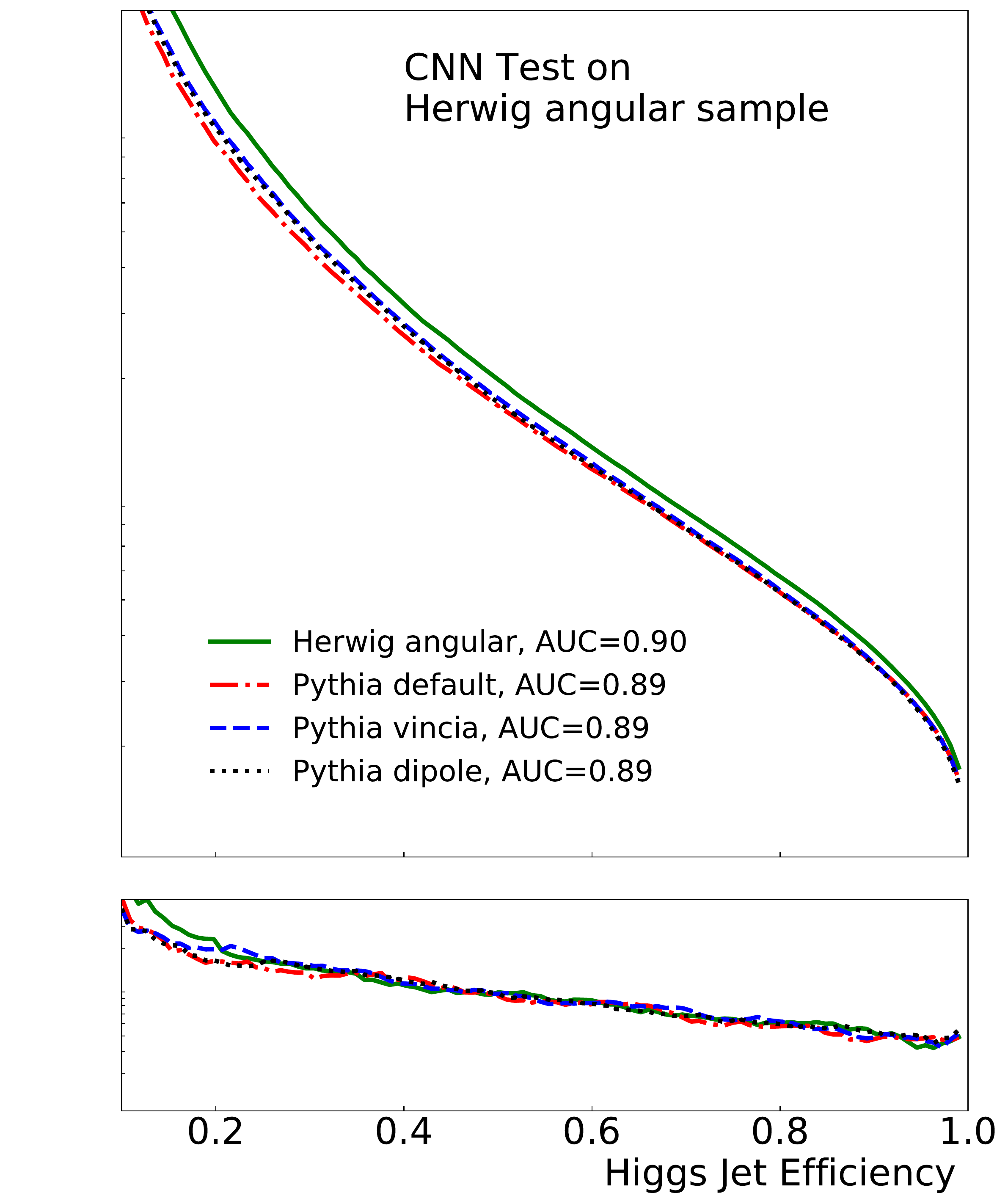}
     \end{subfigure}
     
\caption{The QCD rejection (inverse QCD efficiency) as a function of the Higgs jet efficiency for classifiers applied to  \textsc{Herwig} angular jet from four PSMC algorithms. The bottom panel shows the relative uncertainties.} 
\label{fig:fixed_Herwig_sample}
\end{figure}

Figure \ref{fig:fixed_Herwig_sample} shows four classifiers trained on various simulations and then tested on the same \textsc{Herwig} dataset.  Overall, the CNN has the best performance and the DNN is marginally better than the BDT.  The DNN and BDT are trained on the same features and given the relatively low-dimensionality of the problem, it is unsurprising that the two models have a similar performance.  Overall, the performance is nearly identical for all training sets.  This is even true for the CNN, which has access to low-level substructure information inside the jets.  The insensitivity to the training set is in stark contrast to the sensitivity of the test set, as summarized in detail below.  Additional results can be found in Appendix~\ref{sec:appendix}.



\begin{figure}[h]
\centering
     \begin{subfigure}{0.9\textwidth}
        \centering
        \includegraphics[width=6in]{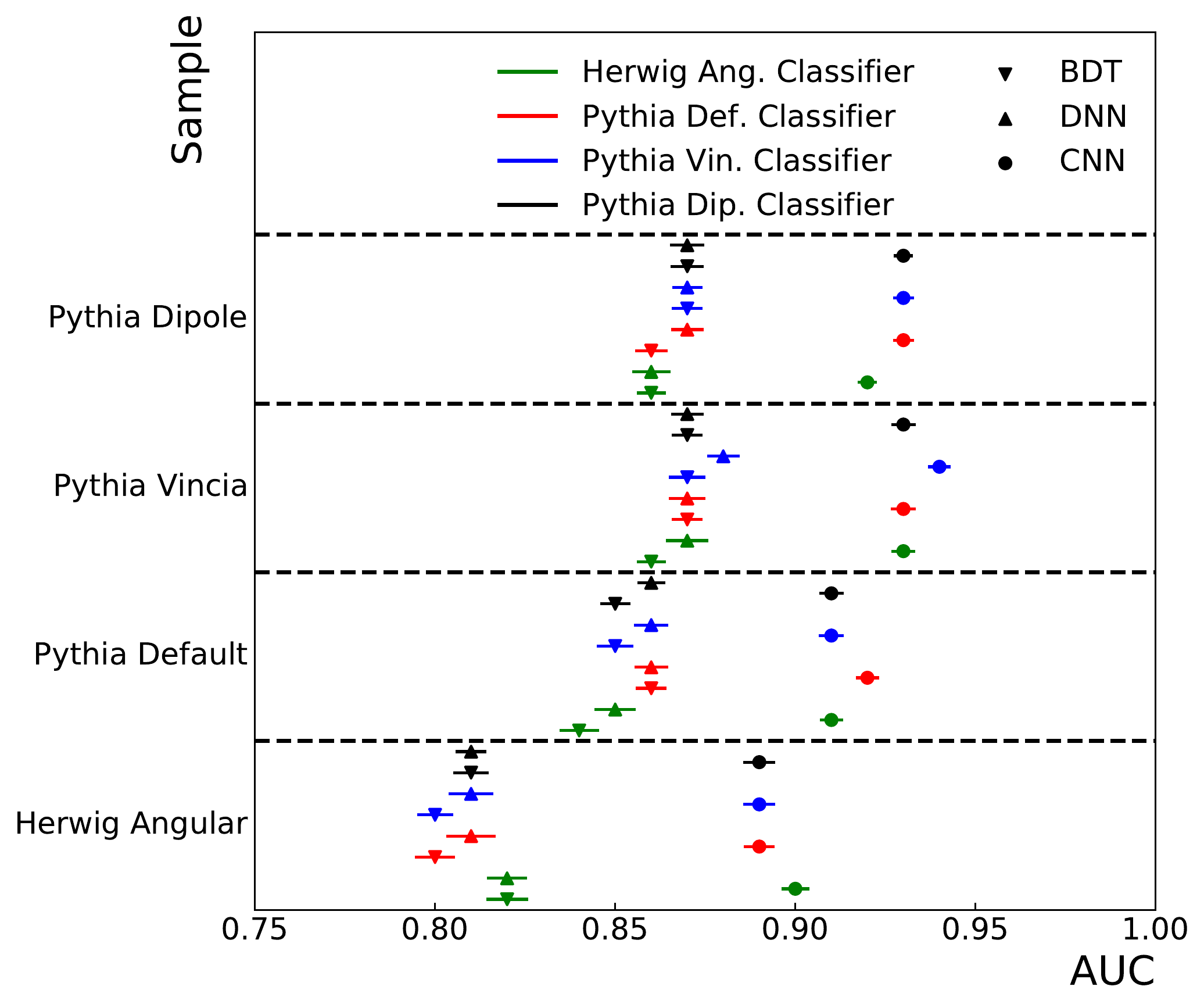}
     \end{subfigure}
\caption{The performance of classifiers as quantified by the AUC when training on a given PSMC (color) and testing on the PSMC specified on the vertical axis.  The symbols represent the type of model (BDT, DNN, CNN). The error bars represent the standard deviation over the $k$ folds.}
\label{fig:auc_summary}
\end{figure}

\begin{figure}[h]
\centering
     \begin{subfigure}{0.9\textwidth}
        \centering
        \includegraphics[width=6in]{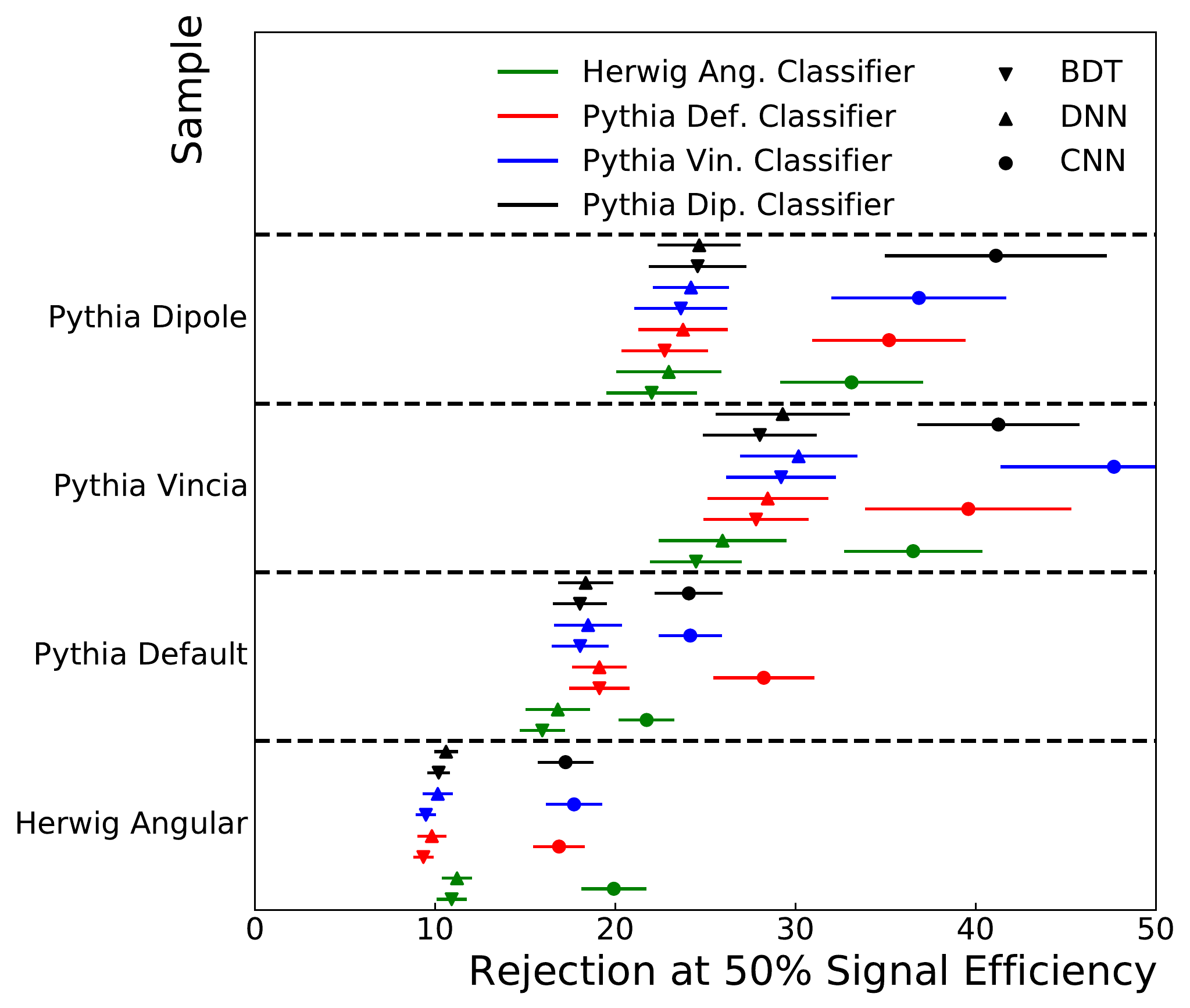}
     \end{subfigure}
\caption{The performance of classifiers as quantified by the rejection at a fixed signal efficiency of 50\% when training on a given PSMC (color) and testing on the PSMC specified on the vertical axis.  The symbols represent the type of model (BDT, DNN, CNN). The error bars represent the standard deviation over the $k$ folds.}
\label{fig:rej_summary}
\end{figure}

\begin{figure}[h]
\centering
     \begin{subfigure}{0.9\textwidth}
        \centering
        \includegraphics[width=6in]{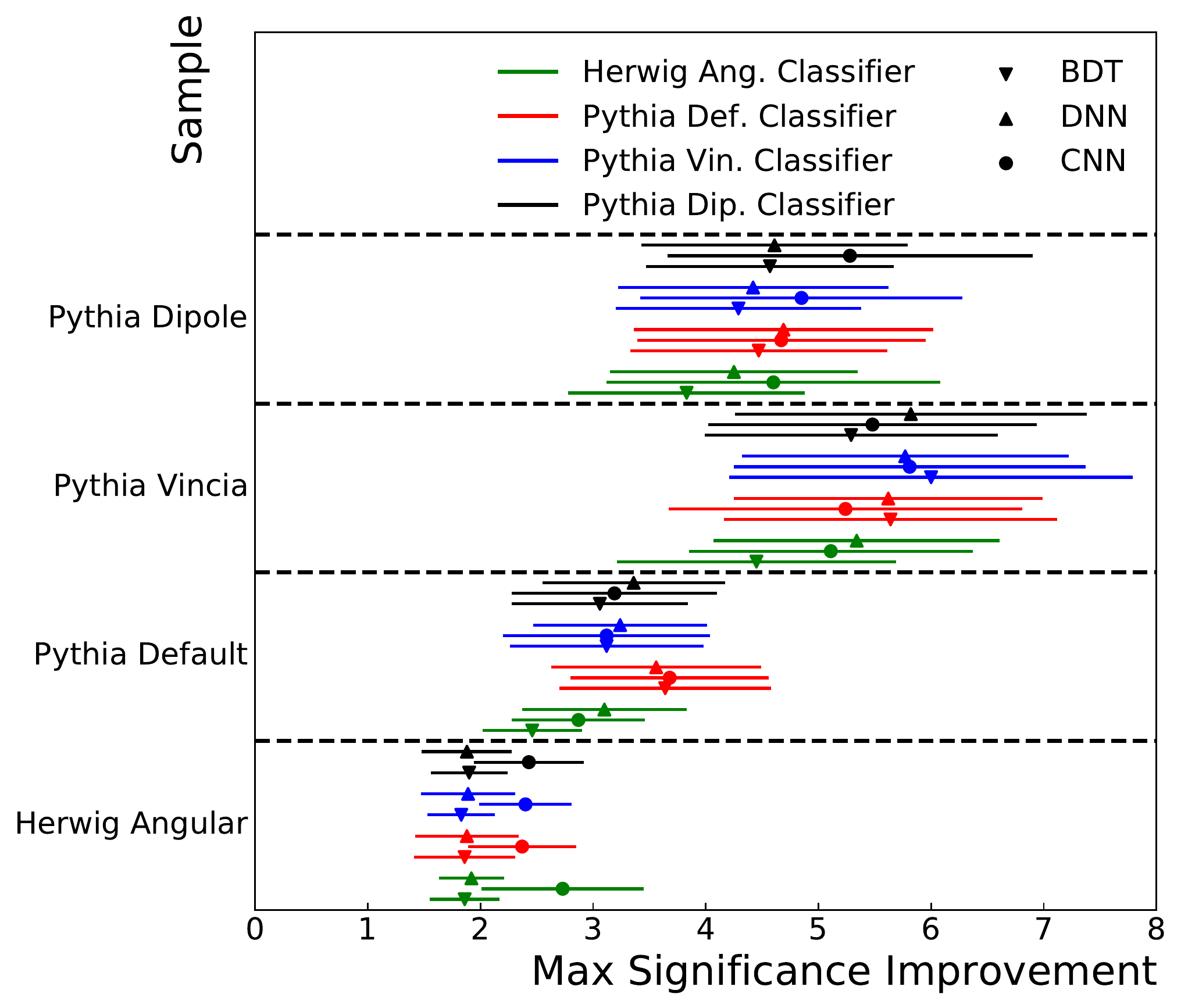}
     \end{subfigure}
\caption{The performance of classifiers as quantified by the maximum significance improvement when training on a given PSMC (color) and testing on the PSMC specified on the vertical axis.  The symbols represent the type of model (BDT, DNN, CNN). The error bars represent the standard deviation over the $k$ folds.}
\label{fig:sig_summary}
\end{figure}

The performance of Fig.~\ref{fig:fixed_Herwig_sample} for all combinations of train and test sets for the three machine learning models are summarized in Fig.~\ref{fig:auc_summary}, \ref{fig:rej_summary}, and~\ref{fig:sig_summary}.  Starting with Fig.~\ref{fig:auc_summary}, we observe that there is a significant spread in performance across test sets (rows).  The difference between Higgs jets and QCD jets is smaller for \textsc{Herwig} compared with \textsc{Pythia} by almost 10\%.  However, the spread in performance for a given test set is about 1\%.  Similar trends are present for the rejection at a fixed efficiency (Fig.~\ref{fig:rej_summary}) and maximum SIC (Fig.~\ref{fig:sig_summary}) plots, albeit with larger sensitivities to the machine learning training.

 \clearpage
\section{Conclusions and Outlook}\label{Conclusions and Outlook}

We have explored the universality of classifiers trained on hadronic jet tagging.  In particular, we have studied the sensitivity of the learned classifier to the Parton Shower Monte Carlo program used during training.  While the modeling of the hadronic structure differs significantly among PSMCs, we find that the actual function learned is nearly independent of the training set. This gives us confidence that a classifier trained on one PSMC and tested on another (or data) will likely still be optimal.  
Although it is not directly a source of uncertainty for physics analysis, this observation has important implications for making the best use of our data.  The classifier universality does not mean that the systematic uncertainty from hadronic modeling is small as bias and optimality are separate concepts (see e.g., Ref.~\cite{Nachman:2019dol}).

The universality not only has important experimental implications, but also motivates further theoretical studies.  As in the quark versus gluon jet example referenced in Sec.~\ref{introduction}, the universality of the classifiers suggests that a theoretical explanation of the classification performance may be attainable as it should be insensitive to the detailed modeling assumptions of a particular PSMC program.  We look forward to studies in this direction.

Uncertainty quantification is a critical component of any analysis at the LHC and this task is particularly challenging for analysis strategies like machine learning that are sensitive to low-level hadronic modeling.  While determining systematic uncertainties on the potential bias of a result from hadronic modeling is still an active area of research and development, we have shown that at least the optimality of machine learning classifiers is relatively insensitive to hadronic modeling.  While we have observed this disconnect between bias and optimality for Higgs jet tagging, we conjecture that this is a generic feature of QCD and it may also be present in other systems at the LHC and beyond.



\acknowledgments

K. Cheung and Y.-L.Chung are supported by the Taiwan MoST with the grant number MOST-110-2112-M-007-017-MY3.   S.-C. Hsu is supported by the U.S. Department of Energy, Office of Science, Office of Early Career Research Program under Award number DE-SC0015971.  B. Nachman is supported by the U.S. Department of Energy, Office of Science under contract DE-AC02-05CH11231.

\clearpage
\appendix
\section{Remaining Results}
\label{sec:appendix}

\begin{table}[h!]
\scriptsize
\begin{center}
\begin{tabular}{c|ccc }
\multicolumn{4}{c}{\textbf{Varied trained classifiers, test on \textsc{Herwig} angular sample}}\\
\hline\hline
\multicolumn{4}{c}{}\\
\multicolumn{4}{c}{\textbf{Metric: Area Under the Curve}}\\
\hline
\diagbox{\textbf{Trained Model}}{\textbf{Classifier Type}} &\textbf{BDT}&\textbf{dense neural network}&\textbf{CNN}\\
\hline
\textbf{Herwig Angular} & $0.82\pm0.0058$ & $0.82\pm0.0056$ & $0.90\pm0.0039$ \\
\textbf{Pythia Default} & $0.80\pm0.0056$ & $0.81\pm0.0069$ & $0.89\pm0.0043$ \\
\textbf{Pythia Vincia} & $0.80\pm0.0050$ & $0.81\pm0.0062$ & $0.89\pm0.0044$ \\
\textbf{Pythia Dipole} & $0.81\pm0.0049$ & $0.81\pm0.0043$ & $0.89\pm0.0044$ \\
\hline
\textbf{Average $\pm$ Std.} & $0.81\pm0.0064$ & $0.81\pm0.0054$ & $0.89\pm0.0047$ \\
\hline\hline
\multicolumn{4}{c}{}\\
\multicolumn{4}{c}{\textbf{Metric: Rejection at 50\% Signal Efficiency}}\\
\hline
\diagbox{\textbf{Trained Model}}{\textbf{Classifier Type}} &\textbf{BDT}&\textbf{dense neural network}&\textbf{CNN}\\
\hline
\textbf{Herwig Angular} &$10.91\pm0.84$ & $11.21\pm0.83$ & $19.91\pm1.81$\\
\textbf{Pythia Default} &$9.34\pm0.57$ & $9.81\pm0.80$ & $16.87\pm1.43$\\
\textbf{Pythia Vincia} &$9.48\pm0.57$  & $10.14\pm0.85$ & $17.70\pm1.56$\\
\textbf{Pythia Dipole} & $10.19\pm0.64$ & $10.60\pm0.66$ & $17.23\pm1.55$\\
\hline
\textbf{Average $\pm$ Std.} &$9.98\pm0.63$ & $10.44\pm0.52$ & $17.93\pm1.18$\\
\hline\hline
\multicolumn{4}{c}{}\\
\multicolumn{4}{c}{\textbf{Metric: Max Significance Improvement}}\\
\hline
\diagbox{\textbf{Trained Model}}{\textbf{Classifier Type}} &\textbf{BDT}&\textbf{dense neural network}&\textbf{CNN}\\
\hline
\textbf{Herwig Angular} & $1.86\pm0.31$ & $1.92\pm0.29$ & $2.73\pm0.72$\\
\textbf{Pythia Default} & $1.86\pm0.45$ & $1.88\pm0.46$ & $2.37\pm0.48$\\
\textbf{Pythia Vincia} &  $1.83\pm0.30$ & $1.89\pm0.42$ & $2.40\pm0.41$\\
\textbf{Pythia Dipole} & $1.90\pm0.34$ & $1.88\pm0.40$  & $2.43\pm0.49$\\
\hline
\textbf{Average $\pm$ Std.} & $1.87\pm0.03$ & $1.89\pm0.02$ & $2.48\pm0.14$\\
\hline\hline

\end{tabular}
\end{center}
\caption{Area under the curve, rejection at 50\% signal efficiency and maximum significance improvement when testing on \textsc{Herwig} for each trained classifier. The last rows are the average and standard deviation over the mean values from the other rows.}
\label{table:fixed_Herwig_sample}
\end{table} 

\begin{figure}[h]
\centering
     \begin{subfigure}{0.3\textwidth}
        \centering
        \includegraphics[width=2.2in]{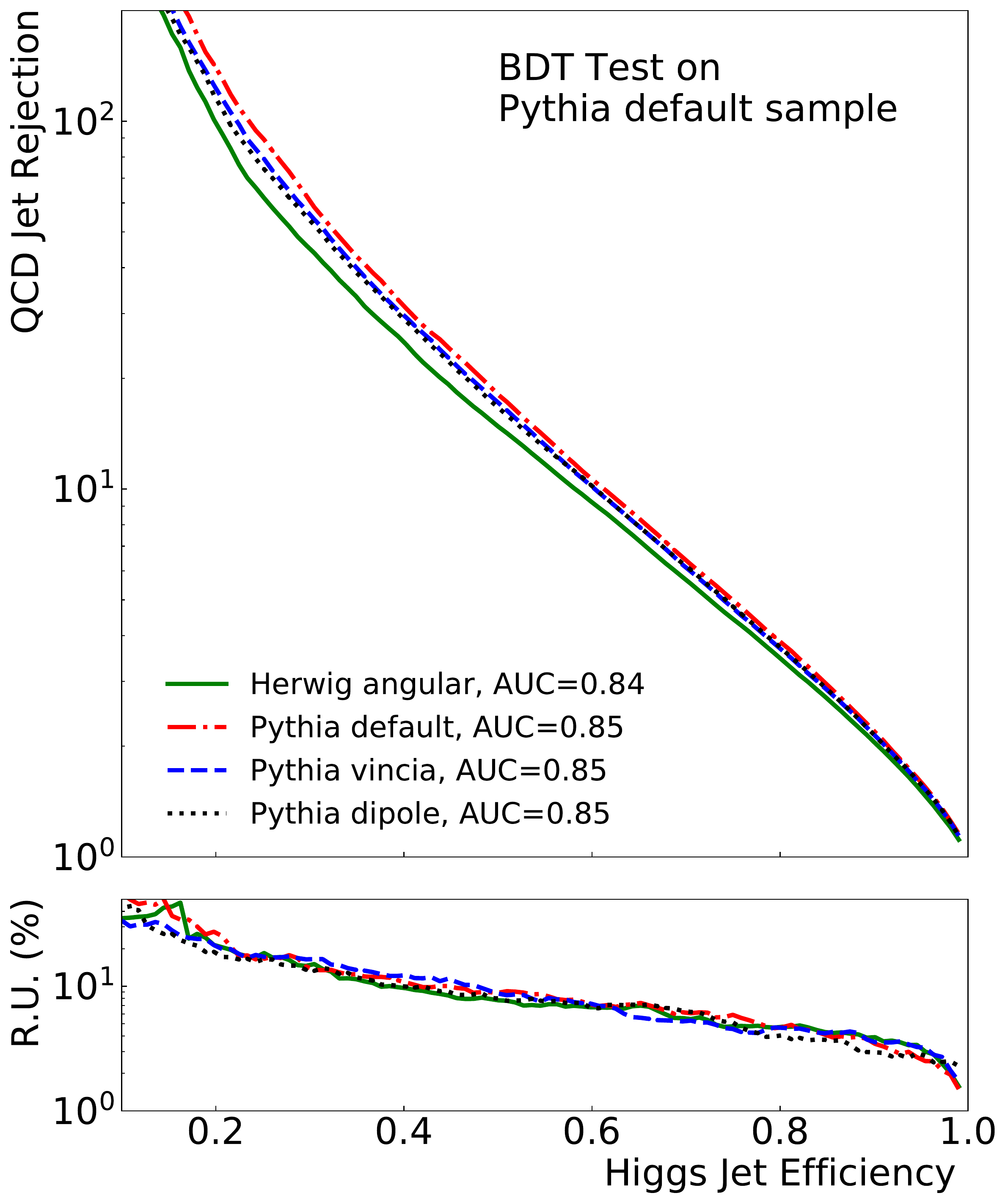}
     \end{subfigure}
     \begin{subfigure}{0.3\textwidth}
        \centering
        \includegraphics[width=2.2in]{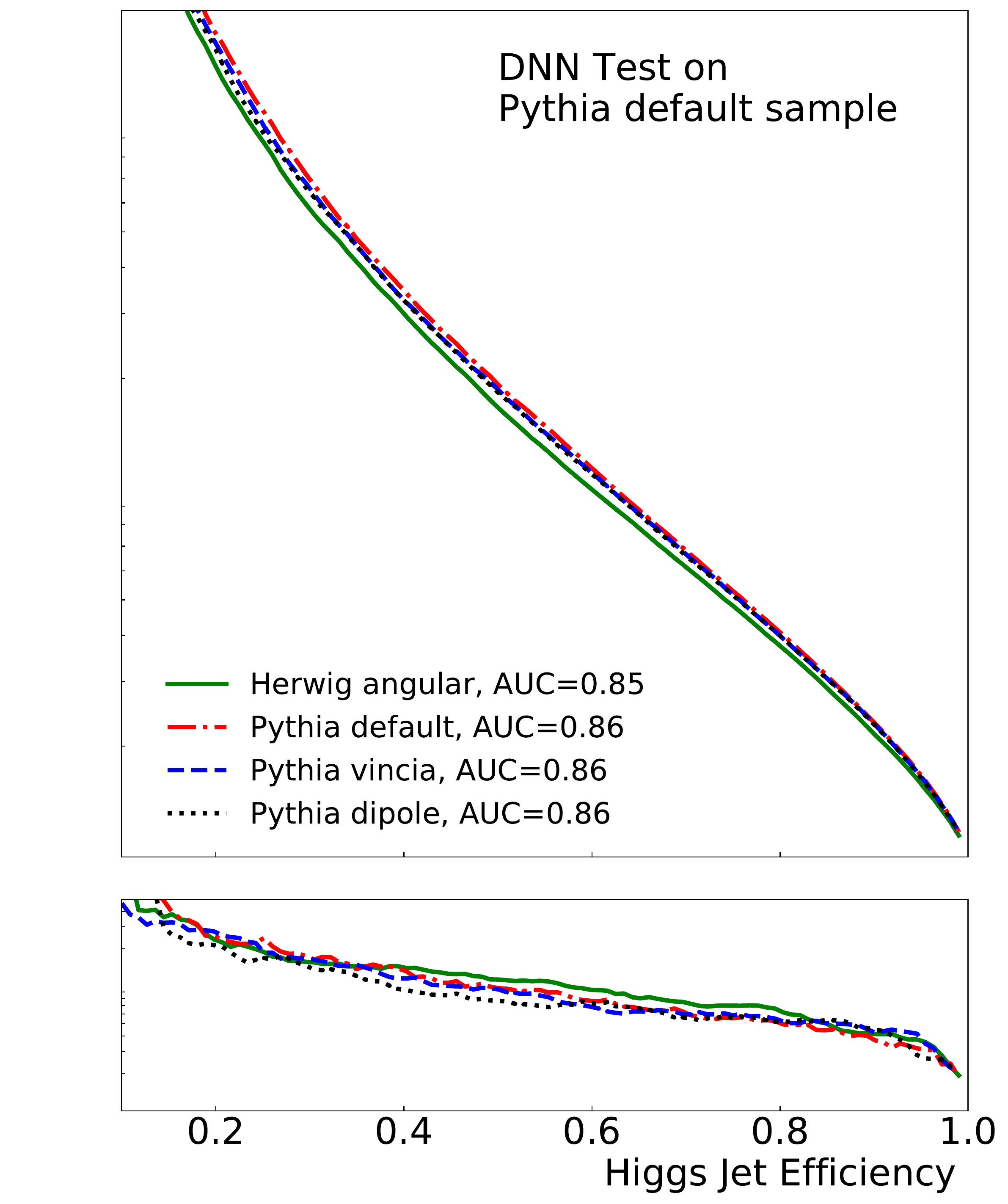}
     \end{subfigure}
     \begin{subfigure}{0.3\textwidth}
        \centering
        \includegraphics[width=2.2in]{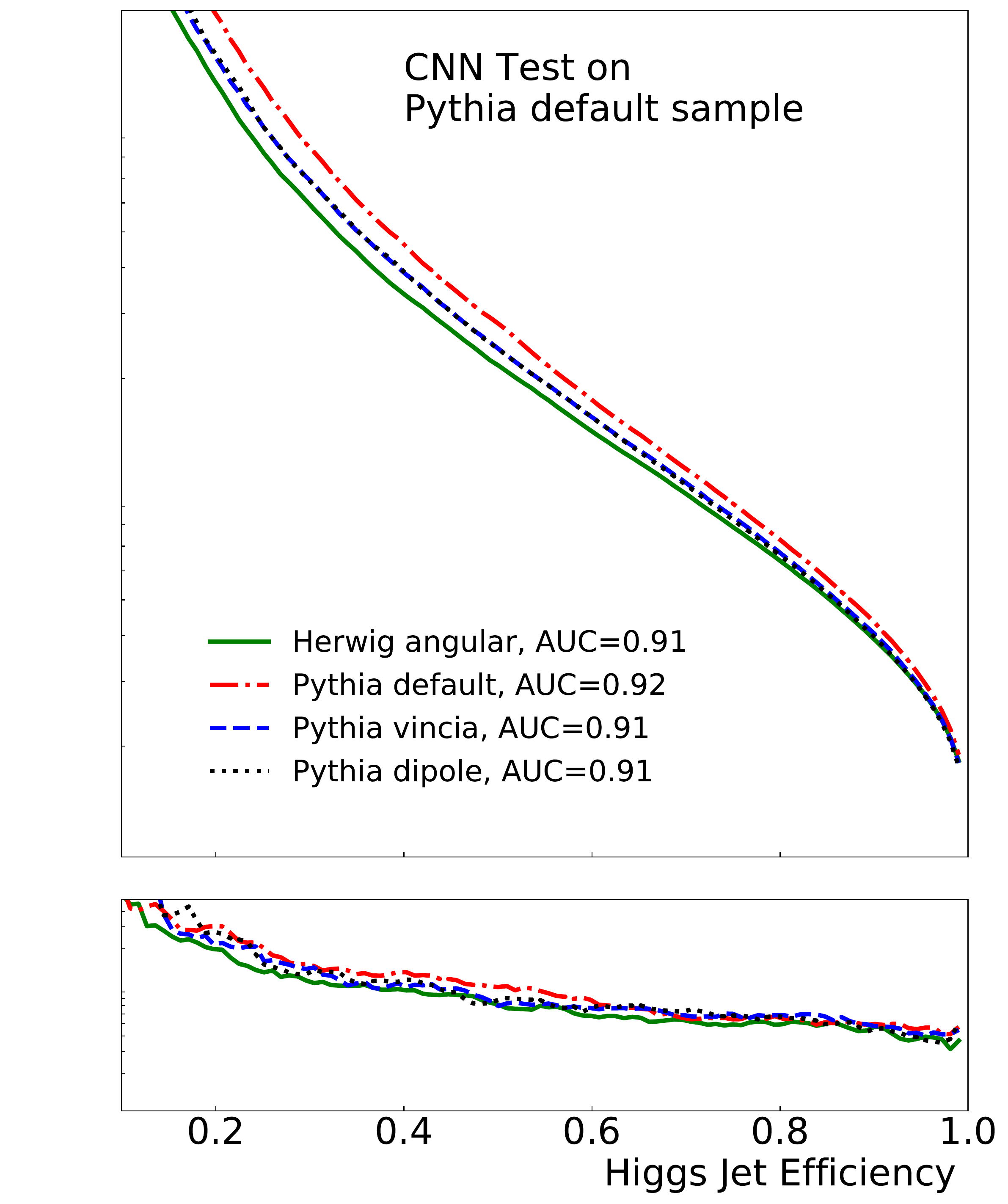}
     \end{subfigure}
\caption{The QCD rejection (inverse QCD efficiency) as a function of the Higgs jet efficiency for classifiers applied to  \textsc{Pythia} default sample from four PSMC algorithms. The bottom panel shows the relative uncertainties.}
\label{fig:fixed_Pythia_def_sample}
\end{figure}

\begin{table}[h!]
\scriptsize
\begin{center}
\begin{tabular}{c|ccc }
\multicolumn{4}{c}{\textbf{Varied trained classifiers, test on \textsc{Pythia} default sample}}\\
\hline\hline
\multicolumn{4}{c}{}\\
\multicolumn{4}{c}{\textbf{Metric: Area Under the Curve}}\\
\hline
\diagbox{\textbf{Trained Model}}{\textbf{Classifier Type}} &\textbf{BDT}&\textbf{dense neural network}&\textbf{CNN}\\
\hline
\textbf{Herwig Angular} & $0.84\pm0.0055$ & $0.85\pm0.0057$ & $0.91\pm0.0032$ \\
\textbf{Pythia Default} & $0.86\pm0.0043$ & $0.86\pm0.0047$ & $0.92\pm0.0032$ \\
\textbf{Pythia Vincia} & $0.85\pm0.0051$ & $0.86\pm0.0048$ & $0.91\pm0.0035$ \\
\textbf{Pythia Dipole} & $0.85\pm0.0042$ & $0.86\pm0.0039$ & $0.91\pm0.0034$ \\
\hline
\textbf{Average $\pm$ Std.} & $0.85\pm0.0053$ & $0.85\pm0.0042$ & $0.91\pm0.0041$ \\
\hline\hline
\multicolumn{4}{c}{}\\
\multicolumn{4}{c}{\textbf{Metric: Rejection at 50\% Signal Efficiency}}\\
\hline
\diagbox{\textbf{Trained Model}}{\textbf{Classifier Type}} &\textbf{BDT}&\textbf{dense neural network}&\textbf{CNN}\\
\hline
\textbf{Herwig Angular} & $15.94\pm1.25$ & $16.80\pm1.78$ & $21.73\pm1.55$ \\
\textbf{Pythia Default} & $19.11\pm1.67$ & $19.11\pm1.51$ & $28.23\pm2.81$ \\
\textbf{Pythia Vincia} & $18.04\pm1.57$ & $18.48\pm1.89$ & $24.15\pm1.75$ \\
\textbf{Pythia Dipole} & $18.03\pm1.51$ & $18.35\pm1.52$ & $24.07\pm1.89$ \\
\hline
\textbf{Average $\pm$ Std.} & $17.78\pm1.15$ & $18.18\pm0.85$ & $24.55\pm2.34$ \\
\hline\hline
\multicolumn{4}{c}{}\\
\multicolumn{4}{c}{\textbf{Metric: Max Significance Improvement}}\\
\hline
\diagbox{\textbf{Trained Model}}{\textbf{Classifier Type}} &\textbf{BDT}&\textbf{dense neural network}&\textbf{CNN}\\
\hline
\textbf{Herwig Angular} & $2.46\pm0.44$ & $3.10\pm0.73$ & $2.87\pm0.59$ \\
\textbf{Pythia Default} & $3.64\pm0.94$ & $3.56\pm0.93$ & $3.68\pm0.88$ \\
\textbf{Pythia Vincia} & $3.12\pm0.86$ & $3.24\pm0.77$ & $3.12\pm0.92$ \\
\textbf{Pythia Dipole} & $3.06\pm0.78$ & $3.36\pm0.81$ & $3.19\pm0.91$ \\
\hline
\textbf{Average $\pm$ Std.} & $3.07\pm0.42$ & $3.31\pm0.17$ & $3.21\pm0.29$ \\
\hline\hline

\end{tabular}
\end{center}
\caption{Area under the curve, rejection at 50\% signal efficiency and maximum significance improvement when testing on \textsc{Pythia} default for each trained classifier. The last rows are the average and standard deviation over the mean values from the other rows.}
\label{table:fixed_Pythia_def_sample}
\end{table} 

\begin{figure}[h]
\centering
     \begin{subfigure}{0.3\textwidth}
        \centering
        \includegraphics[width=2.2in]{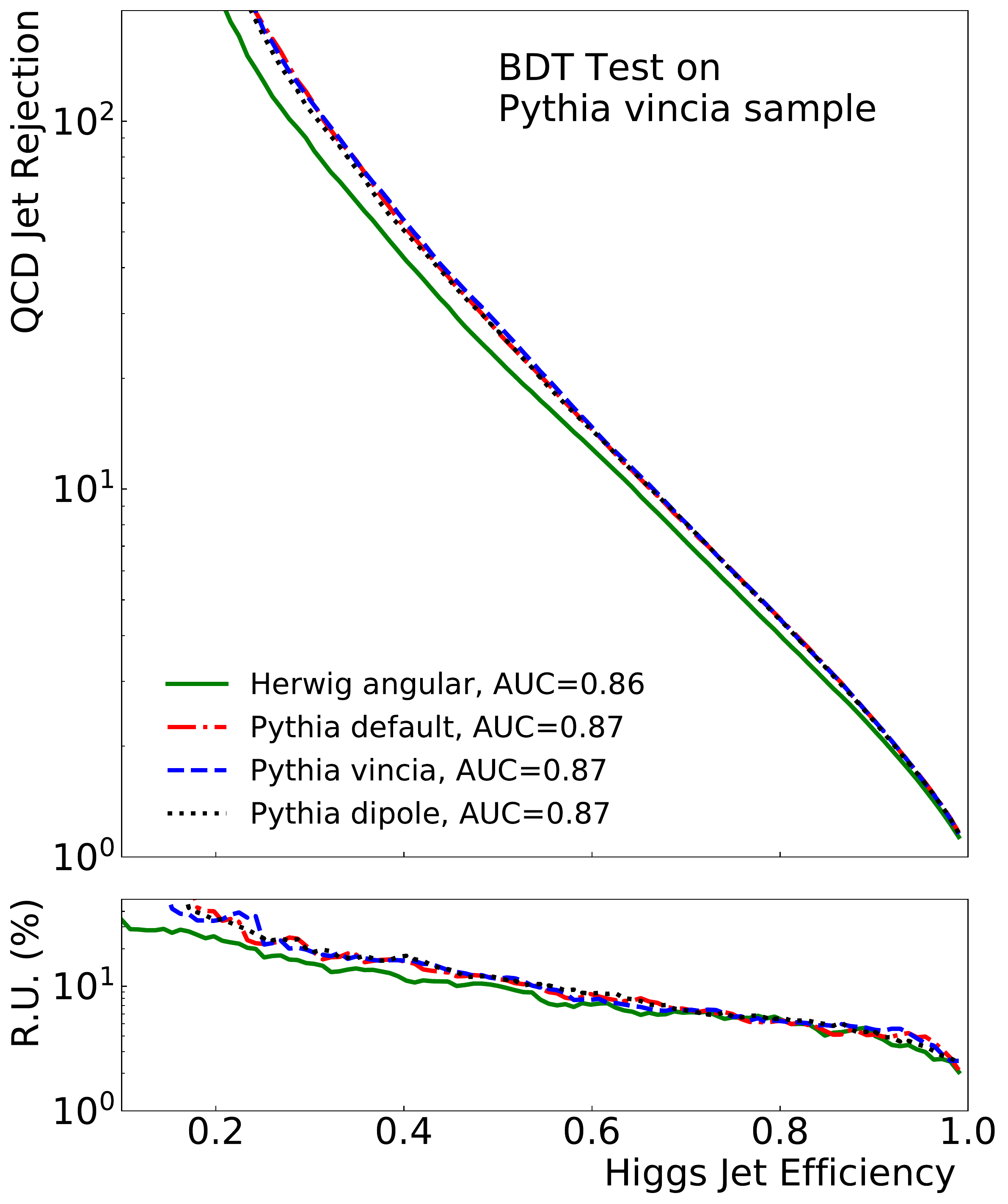}
     \end{subfigure}
     \begin{subfigure}{0.3\textwidth}
        \centering
        \includegraphics[width=2.2in]{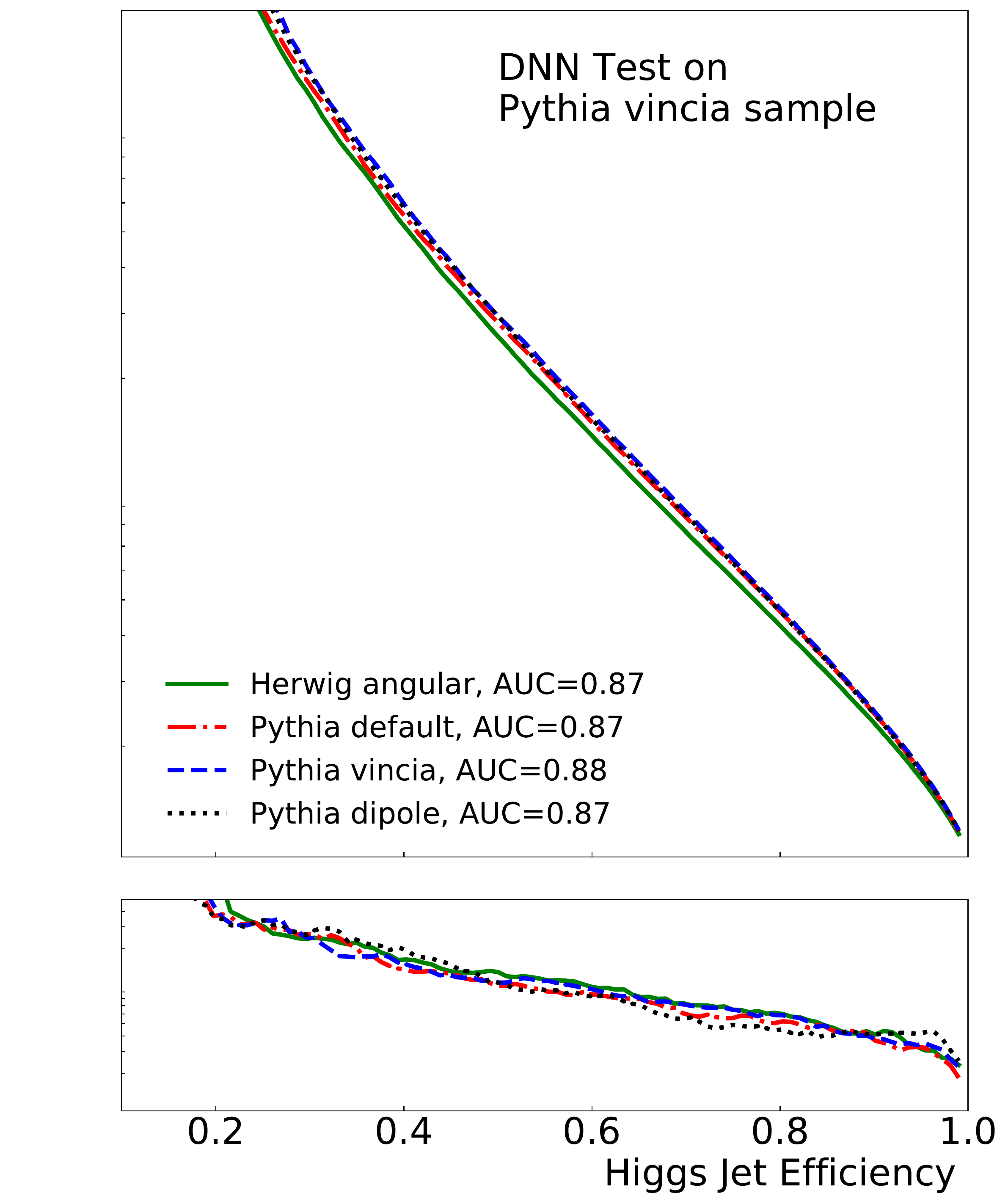}
     \end{subfigure}
     \begin{subfigure}{0.3\textwidth}
        \centering
        \includegraphics[width=2.2in]{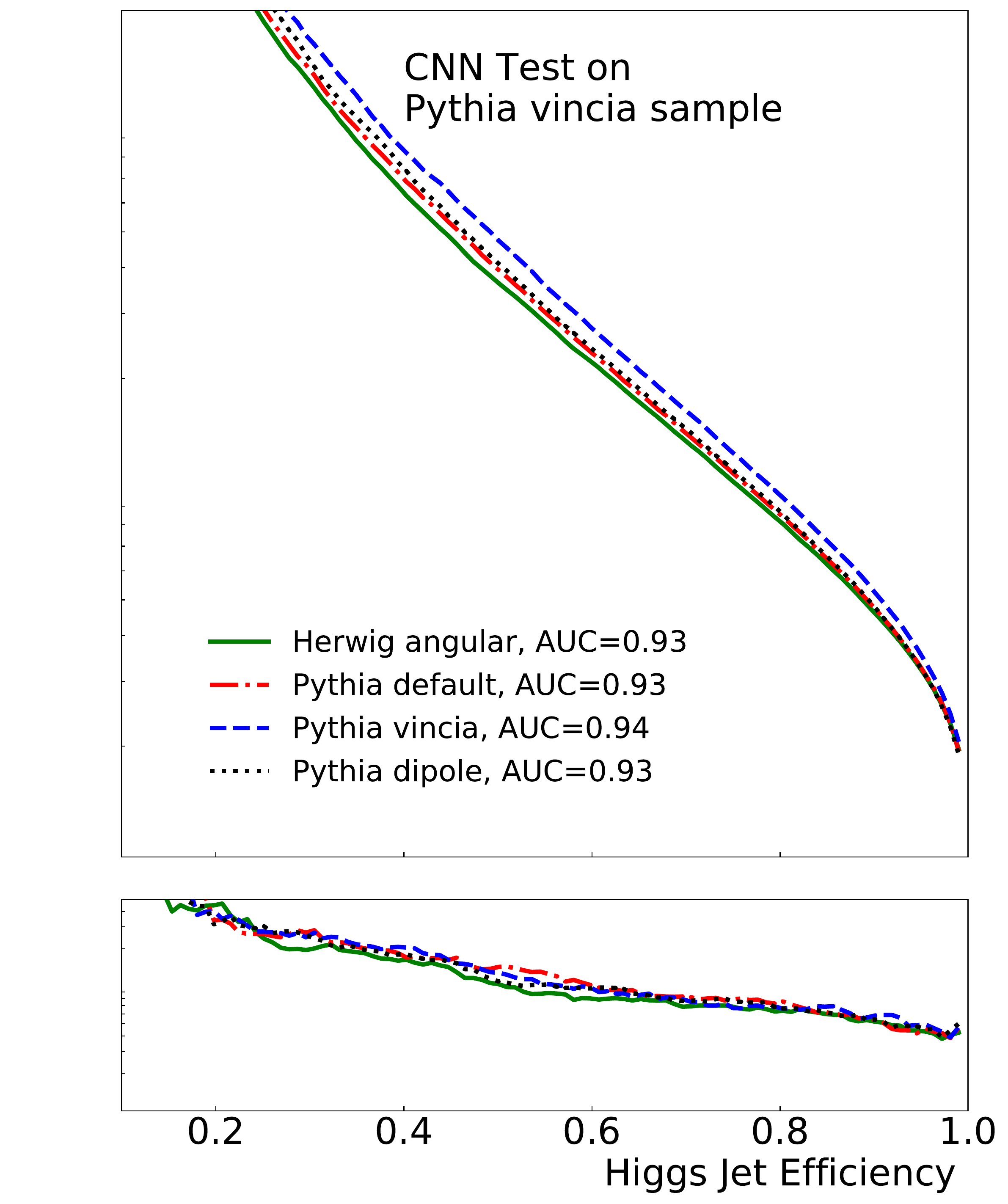}
     \end{subfigure}
\caption{The QCD rejection (inverse QCD efficiency) as a function of the Higgs jet efficiency for classifiers applied to  \textsc{Pythia} VNICIA jet from four PSMC algorithms. The bottom panel shows the relative uncertainties.}
\label{fig:fixed_Pythia_vin_sample}
\end{figure}

\begin{table}[h!]
\scriptsize
\begin{center}
\begin{tabular}{c|ccc }
\multicolumn{4}{c}{\textbf{Varied trained classifiers, test on \textsc{Pythia} VINCIA sample}}\\
\hline\hline
\multicolumn{4}{c}{}\\
\multicolumn{4}{c}{\textbf{Metric: Area Under the Curve}}\\
\hline
\diagbox{\textbf{Trained Model}}{\textbf{Classifier Type}} &\textbf{BDT}&\textbf{dense neural network}&\textbf{CNN}\\
\hline
\textbf{Herwig Angular} & $0.86\pm0.0040$ & $0.87\pm0.0059$ & $0.93\pm0.0033$\\
\textbf{Pythia Default} & $0.87\pm0.0043$ & $0.87\pm0.0051$ & $0.93\pm0.0035$\\
\textbf{Pythia Vincia} & $0.87\pm0.0051$ & $0.88\pm0.0045$  & $0.94\pm0.0031$\\
\textbf{Pythia Dipole} & $0.87\pm0.0043$ & $0.87\pm0.0045$ & $0.93\pm0.0034$\\
\hline
\textbf{Average $\pm$ Std.} & $0.87\pm0.0047$ & $0.87\pm0.0040$ & $0.93\pm0.0035$\\
\hline\hline
\multicolumn{4}{c}{}\\
\multicolumn{4}{c}{\textbf{Metric: Rejection at 50\% Signal Efficiency}}\\
\hline
\diagbox{\textbf{Trained Model}}{\textbf{Classifier Type}} &\textbf{BDT}&\textbf{dense neural network}&\textbf{CNN}\\
\hline
\textbf{Herwig Angular} & $24.47\pm2.55$ & $25.94\pm3.54$ & $36.52\pm3.84$\\
\textbf{Pythia Default} & $27.80\pm2.93$ & $28.45\pm3.35$ & $39.58\pm5.73$\\
\textbf{Pythia Vincia} & $29.19\pm3.05$ & $30.16\pm3.26$ & $47.66\pm6.29$\\
\textbf{Pythia Dipole} & $28.01\pm3.16$ & $29.28\pm3.72$ & $41.25\pm4.51$\\
\hline
\textbf{Average $\pm$ Std.} & $27.37\pm1.76$ & $28.46\pm1.57$ & $41.25\pm4.07$\\
\hline\hline
\multicolumn{4}{c}{}\\
\multicolumn{4}{c}{\textbf{Metric: Max Significance Improvement}}\\
\hline
\diagbox{\textbf{Trained Model}}{\textbf{Classifier Type}} &\textbf{BDT}&\textbf{dense neural network}&\textbf{CNN}\\
\hline
\textbf{Herwig Angular} & $4.45\pm1.24$ & $5.34\pm1.27$ & $5.11\pm1.26$\\
\textbf{Pythia Default} & $5.64\pm1.48$ & $5.62\pm1.37$ & $5.24\pm1.57$\\
\textbf{Pythia Vincia} & $6.00\pm1.79$ & $5.77\pm1.45$ & $5.81\pm1.56$\\
\textbf{Pythia Dipole} & $5.29\pm1.30$ & $5.82\pm1.56$ & $5.48\pm1.46$\\
\hline
\textbf{Average $\pm$ Std.} & $5.34\pm0.58$ & $5.64\pm0.19$ & $5.41\pm0.27$\\
\hline\hline
\end{tabular}
\end{center}
\caption{
Area under the curve, rejection at 50\% signal efficiency and maximum significance improvement when testing on \textsc{Pythia} VINCIA for each trained classifier. The last rows are the average and standard deviation over the mean values from the other rows.}
\label{table:fixed_Pythia_vin_sample}
\end{table} 

\begin{figure}[h]
\centering
     \begin{subfigure}{0.3\textwidth}
        \centering
        \includegraphics[width=2.2in]{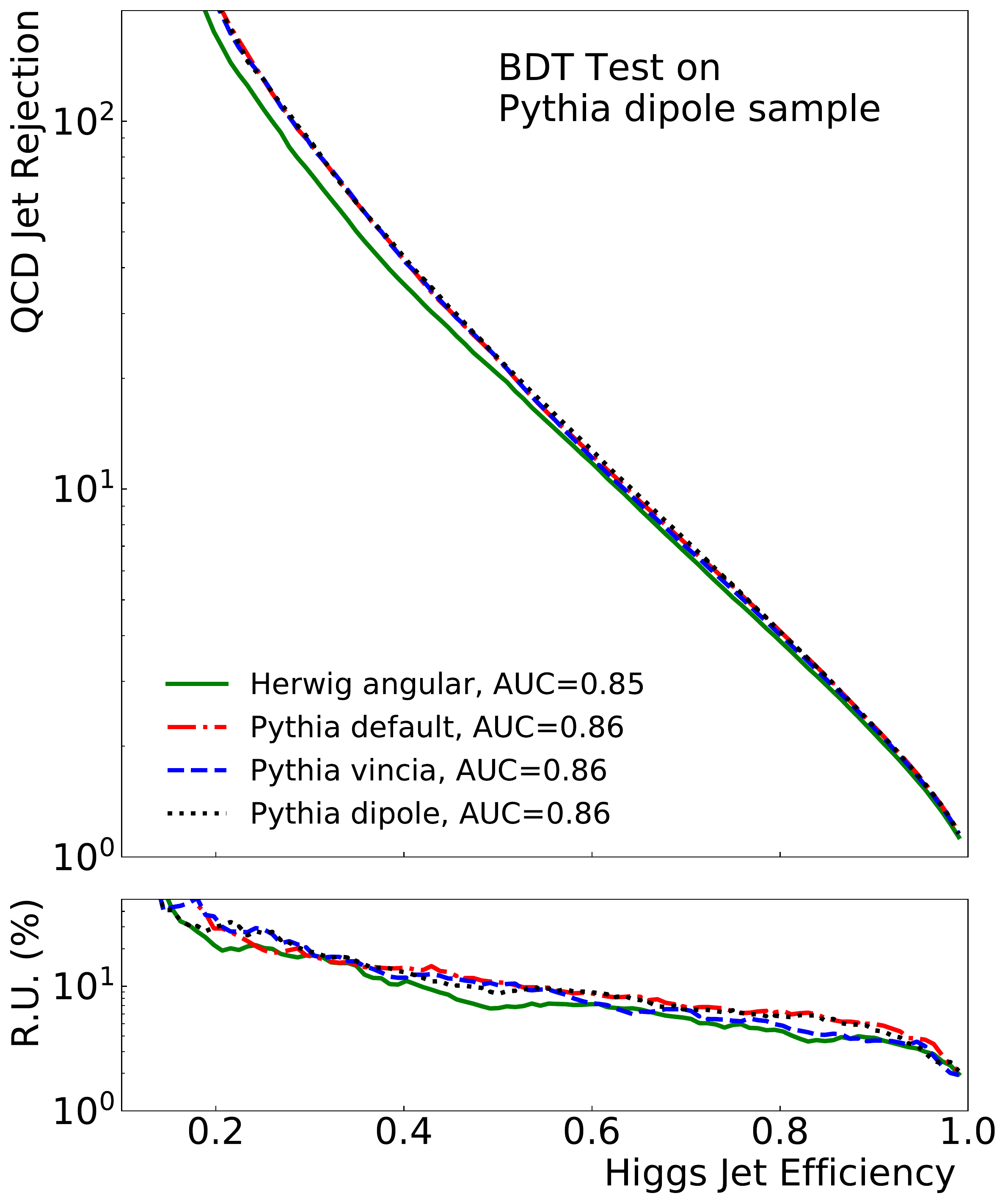}
     \end{subfigure}
     \begin{subfigure}{0.3\textwidth}
        \centering
        \includegraphics[width=2.2in]{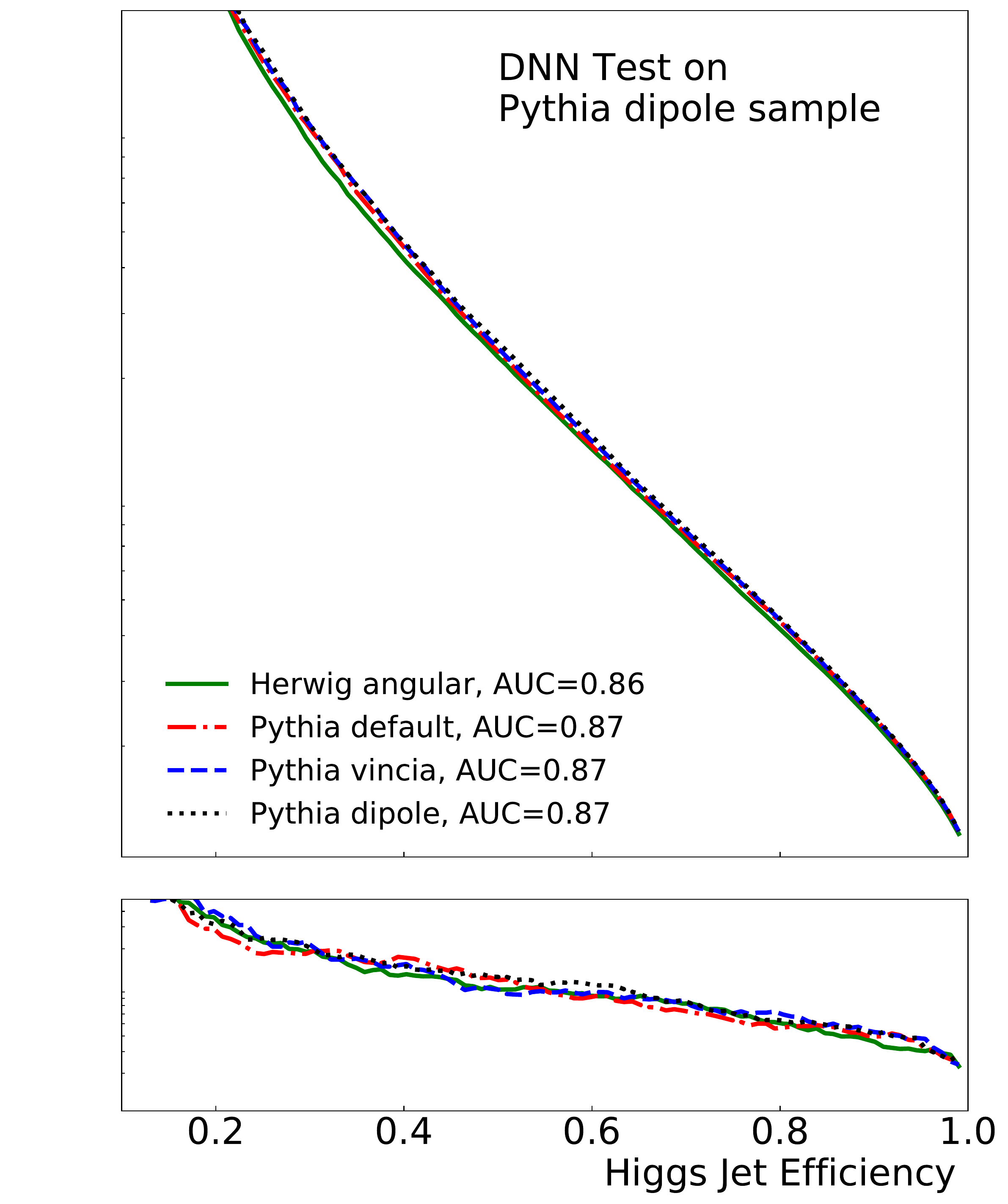}
     \end{subfigure}
     \begin{subfigure}{0.3\textwidth}
        \centering
        \includegraphics[width=2.2in]{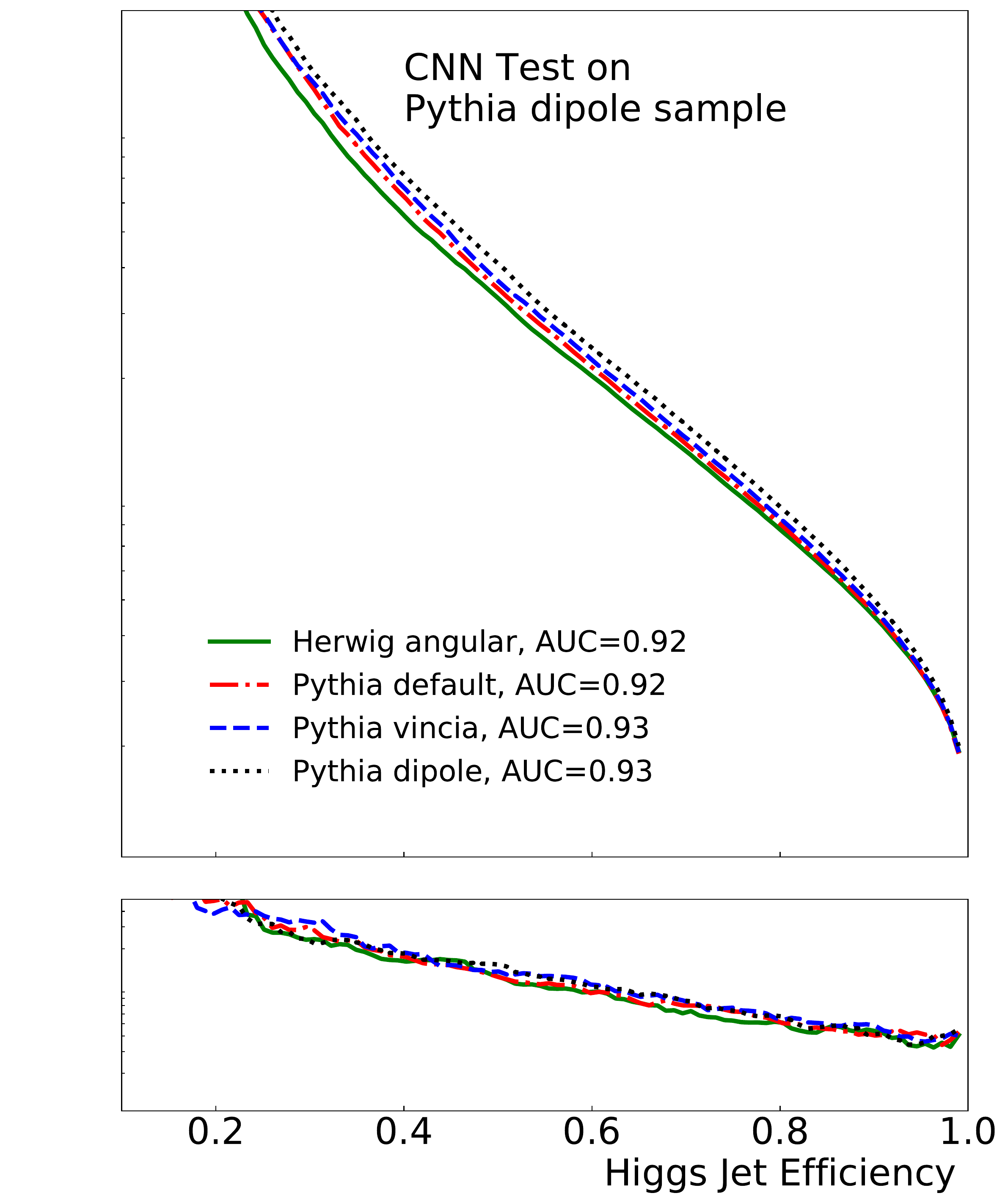}
     \end{subfigure}
\caption{The QCD rejection (inverse QCD efficiency) as a function of the Higgs jet efficiency for classifiers applied to  \textsc{Pythia} dipole jet from four PSMC algorithms. The bottom panel shows the relative uncertainties.}
\label{fig:fixed_Pythia_dip_sample}
\end{figure}

\begin{table}[h!]
\scriptsize
\begin{center}
\begin{tabular}{c|ccc }
\multicolumn{4}{c}{\textbf{Varied trained classifiers, test on \textsc{Pythia} dipole sample}}\\
\hline\hline
\multicolumn{4}{c}{}\\
\multicolumn{4}{c}{\textbf{Metric: Area Under the Curve}}\\
\hline
\diagbox{\textbf{Trained Model}}{\textbf{Classifier Type}} &\textbf{BDT}&\textbf{dense neural network}&\textbf{CNN}\\
\hline
\textbf{Herwig Angular} & $0.86\pm0.0040$ & $0.86\pm0.0053$ & $0.92\pm0.0027$\\
\textbf{Pythia Default} & $0.86\pm0.0045$ & $0.87\pm0.0045$ & $0.92\pm0.0029$\\
\textbf{Pythia Vincia} & $0.87\pm0.0043$ & $0.87\pm0.0042$ & $0.93\pm0.0029$\\
\textbf{Pythia Dipole} & $0.87\pm0.0046$ & $0.87\pm0.0048$ & $0.93\pm0.0027$\\
\hline
\textbf{Average $\pm$ Std.} & $0.86\pm0.0029$ & $0.87\pm0.0025$ & $0.93\pm0.0030$\\
\hline\hline
\multicolumn{4}{c}{}\\
\multicolumn{4}{c}{\textbf{Metric: Rejection at 50\% Signal Efficiency}}\\
\hline
\diagbox{\textbf{Trained Model}}{\textbf{Classifier Type}} &\textbf{BDT}&\textbf{dense neural network}&\textbf{CNN}\\
\hline
\textbf{Herwig Angular} & $22.01\pm2.52$ & $22.96\pm2.91$ & $33.10\pm3.97$\\
\textbf{Pythia Default} & $22.73\pm2.40$ & $23.75\pm2.48$ & $35.18\pm4.26$\\
\textbf{Pythia Vincia} & $23.63\pm2.58$ & $24.19\pm2.12$ & $36.84\pm4.86$\\
\textbf{Pythia Dipole} & $24.56\pm2.72$ & $24.65\pm2.31$ & $41.11\pm6.17$\\
\hline
\textbf{Average $\pm$ Std.} & $23.23\pm0.96$ & $23.89\pm0.62$ & $36.56\pm2.94$\\
\hline\hline
\multicolumn{4}{c}{}\\
\multicolumn{4}{c}{\textbf{Metric: Max Significance Improvement}}\\
\hline
\diagbox{\textbf{Trained Model}}{\textbf{Classifier Type}} &\textbf{BDT}&\textbf{dense neural network}&\textbf{CNN}\\
\hline
\textbf{Herwig Angular} & $3.83\pm1.05$ & $4.25\pm1.10$ & $4.60\pm1.48$\\
\textbf{Pythia Default} & $4.47\pm1.14$ & $4.69\pm1.33$ & $4.67\pm1.28$\\
\textbf{Pythia Vincia} & $4.29\pm1.09$ & $4.42\pm1.20$ & $4.85\pm1.43$\\
\textbf{Pythia Dipole} & $4.57\pm1.10$ & $4.61\pm1.18$ & $5.28\pm1.62$\\
\hline
\textbf{Average $\pm$ Std.} & $4.29\pm0.28$ & $4.49\pm0.17$ & $4.85\pm0.26$\\
\hline\hline
\end{tabular}
\end{center}
\caption{
Area under the curve, rejection at 50\% signal efficiency and maximum significance improvement when testing on \textsc{Pythia} dipole for each trained classifier. The last rows are the average and standard deviation over the mean values from the other rows.}
\label{table:fixed_Pythia_dip_sample}
\end{table} 

\begin{table}[h!]
\scriptsize
\begin{center}
\begin{tabular}{c|cccc }
\multicolumn{5}{c}{\textbf{Varied trained BDT models test on fixed sample}}\\
\hline\hline
\multicolumn{5}{c}{}\\
\multicolumn{5}{c}{\textbf{metric: area under the curve}}\\
\hline
\diagbox{\textbf{Trained Model}}{\textbf{Fixed Sample}} &\textbf{Herwig Angular}&\textbf{Pythia Default}&\textbf{Pythia Vincia}&\textbf{Pythia Dipole}\\
\hline
\textbf{Herwig Angular} & $0.82\pm0.0058$ & $0.84\pm0.0055$ & $0.86\pm0.0040$ & $0.86\pm0.0040$\\
\textbf{Pythia Default} & $0.80\pm0.0056$ & $0.86\pm0.0043$ & $0.87\pm0.0043$ & $0.86\pm0.0045$\\
\textbf{Pythia Vincia} & $0.80\pm0.0050$ & $0.85\pm0.0051$ & $0.87\pm0.0051$ & $0.87\pm0.0043$\\
\textbf{Pythia Dipole} & $0.81\pm0.0049$ & $0.85\pm0.0042$ & $0.87\pm0.0043$ & $0.87\pm0.0046$\\
\hline
\textbf{Average $\pm$ Std.} & $0.81\pm0.0064$ & $0.85\pm0.0053$ & $0.87\pm0.0047$ & $0.86\pm0.0029$\\
\hline\hline
\multicolumn{5}{c}{}\\
\multicolumn{5}{c}{\textbf{metric: rejection at 50\% signal efficiency}}\\
\hline
\diagbox{\textbf{Trained Model}}{\textbf{Fixed Sample}} &\textbf{Herwig Angular}&\textbf{Pythia Default}&\textbf{Pythia Vincia}&\textbf{Pythia Dipole}\\
\hline
\textbf{Herwig Angular} & $10.91\pm0.84$ & $15.94\pm1.25$ & $24.47\pm2.55$ & $22.01\pm2.52$\\
\textbf{Pythia Default} & $9.34\pm0.57$ & $19.11\pm1.67$ & $27.80\pm2.93$ & $22.73\pm2.40$\\
\textbf{Pythia Vincia} & $9.48\pm0.57$ & $18.04\pm1.57$ & $29.19\pm3.05$ & $23.63\pm2.58$\\
\textbf{Pythia Dipole} & $10.19\pm0.64$ & $18.03\pm1.51$ & $28.01\pm3.16$ & $24.56\pm2.72$\\
\hline
\textbf{Average $\pm$ Std.} & $9.98\pm0.63$ & $17.78\pm1.15$ & $27.37\pm1.76$ & $23.23\pm0.96$\\
\hline\hline
\multicolumn{5}{c}{}\\
\multicolumn{5}{c}{\textbf{metric: Max Significance Improvement}}\\
\hline
\diagbox{\textbf{Trained Model}}{\textbf{Fixed Sample}} &\textbf{Herwig Angular}&\textbf{Pythia Default}&\textbf{Pythia Vincia}&\textbf{Pythia Dipole}\\
\hline
\textbf{Herwig Angular} & $1.86\pm0.31$ & $2.46\pm0.44$ & $4.45\pm1.24$ & $3.83\pm1.05$\\
\textbf{Pythia Default} & $1.86\pm0.45$ & $3.64\pm0.94$ & $5.64\pm1.48$ & $4.47\pm1.14$\\
\textbf{Pythia Vincia} & $1.83\pm0.30$ & $3.12\pm0.86$ & $6.00\pm1.79$ & $4.29\pm1.09$\\
\textbf{Pythia Dipole} & $1.90\pm0.34$ & $3.06\pm0.78$ & $5.29\pm1.30$ & $4.57\pm1.10$\\
\hline
\textbf{Average $\pm$ Std.} & $1.87\pm0.03$ & $3.07\pm0.42$ & $5.34\pm0.58$ & $4.29\pm0.28$\\
\hline\hline

\end{tabular}
\end{center}
\caption{Area under the curve, rejection at 50\% signal efficiency and maximum significance improvement for the BDT model. The last rows are the average and standard deviation over the mean values from the other rows.}
\label{table:fixed_data_table_BDT}
\end{table} 

\begin{table}[h!]
\scriptsize
\begin{center}
\begin{tabular}{c|cccc }
\multicolumn{5}{c}{\textbf{Varied trained DNN models test on fixed sample}}\\
\hline\hline
\multicolumn{5}{c}{}\\
\multicolumn{5}{c}{\textbf{metric: area under the curve}}\\
\hline
\diagbox{\textbf{Trained Model}}{\textbf{Fixed Sample}} &\textbf{Herwig Angular}&\textbf{Pythia Default}&\textbf{Pythia Vincia}&\textbf{Pythia Dipole}\\
\hline
\textbf{Herwig Angular} & $0.82\pm0.0056$ & $0.85\pm0.0057$ & $0.87\pm0.0059$ & $0.86\pm0.0053$\\
\textbf{Pythia Default} & $0.81\pm0.0069$ & $0.86\pm0.0047$ & $0.87\pm0.0051$ & $0.87\pm0.0045$\\
\textbf{Pythia Vincia} & $0.81\pm0.0062$ & $0.86\pm0.0048$ & $0.88\pm0.0045$ & $0.87\pm0.0042$\\
\textbf{Pythia Dipole} & $0.81\pm0.0043$ & $0.86\pm0.0039$ & $0.87\pm0.0045$ & $0.87\pm0.0048$\\
\hline
\textbf{Average $\pm$ Std.} & $0.81\pm0.0054$ & $0.85\pm0.0042$ & $0.87\pm0.0040$ & $0.87\pm0.0025$\\
\hline\hline
\multicolumn{5}{c}{}\\
\multicolumn{5}{c}{\textbf{metric: rejection at 50\% signal efficiency}}\\
\hline
\diagbox{\textbf{Trained Model}}{\textbf{Fixed Sample}} &\textbf{Herwig Angular}&\textbf{Pythia Default}&\textbf{Pythia Vincia}&\textbf{Pythia Dipole}\\
\hline
\textbf{Herwig Angular} & $11.21\pm0.83$ & $16.80\pm1.78$ & $25.94\pm3.54$ & $22.96\pm2.91$\\
\textbf{Pythia Default} & $9.81\pm0.80$ & $19.11\pm1.51$ & $28.45\pm3.35$ & $23.75\pm2.48$\\
\textbf{Pythia Vincia} & $10.14\pm0.85$ & $18.48\pm1.89$ & $30.16\pm3.26$ & $24.19\pm2.12$\\
\textbf{Pythia Dipole} & $10.60\pm0.66$ & $18.35\pm1.52$ & $29.28\pm3.72$ & $24.65\pm2.31$\\
\hline
\textbf{Average $\pm$ Std.} & $10.44\pm0.52$ & $18.18\pm0.85$ & $28.46\pm1.57$ & $23.89\pm0.62$\\
\hline\hline
\multicolumn{5}{c}{}\\
\multicolumn{5}{c}{\textbf{metric: Max Significance Improvement}}\\
\hline
\diagbox{\textbf{Trained Model}}{\textbf{Fixed Sample}} &\textbf{Herwig Angular}&\textbf{Pythia Default}&\textbf{Pythia Vincia}&\textbf{Pythia Dipole}\\
\hline
\textbf{Herwig Angular} & $1.92\pm0.29$ & $3.10\pm0.73$ & $5.34\pm1.27$ & $4.25\pm1.10$\\
\textbf{Pythia Default} & $1.88\pm0.46$ & $3.56\pm0.93$ & $5.62\pm1.37$ & $4.69\pm1.33$\\
\textbf{Pythia Vincia} & $1.89\pm0.42$ & $3.24\pm0.77$ & $5.77\pm1.45$ & $4.42\pm1.20$\\
\textbf{Pythia Dipole} & $1.88\pm0.40$ & $3.36\pm0.81$ & $5.82\pm1.56$ & $4.61\pm1.18$\\
\hline
\textbf{Average $\pm$ Std.} & $1.89\pm0.02$ & $3.31\pm0.17$ & $5.64\pm0.19$ & $4.49\pm0.17$\\
\hline\hline
\end{tabular}
\end{center}
\caption{Area under the curve, rejection at 50\% signal efficiency and maximum significance improvement for the DNN model. The last rows are the average and standard deviation over the mean values from the other rows.}
\label{table:fixed_data_table_DNN}
\end{table} 

\begin{table}[h!]
\scriptsize
\begin{center}
\begin{tabular}{c|cccc }
\multicolumn{5}{c}{\textbf{Varied trained CNN models test on fixed sample}}\\
\hline\hline
\multicolumn{5}{c}{}\\
\multicolumn{5}{c}{\textbf{metric: area under the curve}}\\
\hline
\diagbox{\textbf{Trained Model}}{\textbf{Fixed Sample}} &\textbf{Herwig Angular}&\textbf{Pythia Default}&\textbf{Pythia Vincia}&\textbf{Pythia Dipole}\\
\hline
\textbf{Herwig Angular} & $0.90\pm0.0039$ & $0.91\pm0.0032$ & $0.93\pm0.0033$ & $0.92\pm0.0027$\\
\textbf{Pythia Default} & $0.89\pm0.0043$ & $0.92\pm0.0032$ & $0.93\pm0.0035$ & $0.92\pm0.0029$\\
\textbf{Pythia Vincia} & $0.89\pm0.0044$ & $0.91\pm0.0035$ & $0.94\pm0.0031$ & $0.93\pm0.0029$\\
\textbf{Pythia Dipole} & $0.89\pm0.0044$ & $0.91\pm0.0034$ & $0.93\pm0.0034$ & $0.93\pm0.0027$
\\
\hline
\textbf{Average $\pm$ Std.} & $0.89\pm0.0047$ & $0.91\pm0.0041$ & $0.93\pm0.0035$ & $0.93\pm0.0030$\\
\hline\hline
\multicolumn{5}{c}{}\\
\multicolumn{5}{c}{\textbf{metric: rejection at 50\% signal efficiency}}\\
\hline
\diagbox{\textbf{Trained Model}}{\textbf{Fixed Sample}} &\textbf{Herwig Angular}&\textbf{Pythia Default}&\textbf{Pythia Vincia}&\textbf{Pythia Dipole}\\
\hline
\textbf{Herwig Angular} & $19.91\pm1.81$ & $21.73\pm1.55$ & $36.52\pm3.84$ & $33.10\pm3.97$\\
\textbf{Pythia Default} & $16.87\pm1.43$ & $28.23\pm2.81$ & $39.58\pm5.73$ & $35.18\pm4.26$\\
\textbf{Pythia Vincia} & $17.70\pm1.56$ & $24.15\pm1.75$ & $47.66\pm6.29$ & $36.84\pm4.86$\\
\textbf{Pythia Dipole} & $17.23\pm1.55$ & $24.07\pm1.89$ & $41.25\pm4.51$ & $41.11\pm6.17$\\
\hline
\textbf{Average $\pm$ Std.} & $17.93\pm1.18$ & $24.55\pm2.34$ & $41.25\pm4.07$ & $36.56\pm2.94$\\
\hline\hline
\multicolumn{5}{c}{}\\
\multicolumn{5}{c}{\textbf{metric: Max Significance Improvement}}\\
\hline
\diagbox{\textbf{Trained Model}}{\textbf{Fixed Sample}} &\textbf{Herwig Angular}&\textbf{Pythia Default}&\textbf{Pythia Vincia}&\textbf{Pythia Dipole}\\
\hline
\textbf{Herwig Angular} & $2.73\pm0.72$ & $2.87\pm0.59$ & $5.11\pm1.26$ & $4.60\pm1.48$\\
\textbf{Pythia Default} & $2.37\pm0.48$ & $3.68\pm0.88$ & $5.24\pm1.57$ & $4.67\pm1.28$\\
\textbf{Pythia Vincia} & $2.40\pm0.41$ & $3.12\pm0.92$ & $5.81\pm1.56$ & $4.85\pm1.43$\\
\textbf{Pythia Dipole} & $2.43\pm0.49$ & $3.19\pm0.91$ & $5.48\pm1.46$ & $5.28\pm1.62$\\
\hline
\textbf{Average $\pm$ Std.} & $2.48\pm0.14$ & $3.21\pm0.29$ & $5.41\pm0.27$ & $4.85\pm0.26$\\
\hline\hline

\end{tabular}
\end{center}
\caption{Area under the curve, rejection at 50\% signal efficiency and maximum significance improvement for the CNN model. The last rows are the average and standard deviation over the mean values from the other rows.}
\label{table:fixed_data_table_CNN}
\end{table} 

\clearpage
\bibliographystyle{jhep}
\bibliography{references,HEPML}

\end{document}